\documentclass[british,english,authoryear,numbers]{article}
\usepackage[T1]{fontenc}
\usepackage[latin9]{inputenc}
\usepackage{geometry}
\usepackage{graphicx}
\usepackage[authoryear]{natbib}
\usepackage{nomencl}

\providecommand{\makenomenclature}{\makeglossary}
\makenomenclature

\makeatletter

\providecommand{\tabularnewline}{\\}

\newcommand{\lyxaddress}[1]{
\par {\raggedright #1
\vspace{1.4em}
\noindent\par}
}

\usepackage[latin9]{inputenc}
\usepackage{gensymb}

\@ifundefined{showcaptionsetup}{}{%
 \PassOptionsToPackage{caption=false}{subfig}}
\usepackage{subfig}
\makeatother

\usepackage{babel}
\begin{document}

\title{A multiple model and observation 10m wind climatology for the Gulf
of California}

\author{Markus Gross}

\maketitle

\lyxaddress{CICESE (Centro de Investigación Científica y de Educación Superior
de Ensenada), Departamento de Oceanografía Física, Carretera Ensenada-Tijuana
3918, Ensenada BC 22860, MEXICO, mgross@cicese.mx}
\begin{abstract}
Seven data sources are \foreignlanguage{british}{analysed} and combined
to form a surface wind speed climatology for the Gulf of California.
\end{abstract}

\section*{Dataset}
\begin{enumerate}
\item name of dataset: 10m Wind Speed Climatology for the Gulf of California\\
data centre: figshare\\
DOI:10.6084/m9.figshare.4037235\\
Identifier: xxxxx \\
Creator: Markus Gross \\
Title: xxxxxx \\
Publisher: xxxxxx \\
Publication year: 2016 \\
Resource type: netcdf and ASCII files \\
Version: 3.0
\end{enumerate}

\section{Introduction}

Climatologically correct data for $10$m wind speeds are important
for a variety of applications, such as forcing regional ocean models,
which then in turn can be used to \foreignlanguage{british}{analyse}
tracer transport, biological activity, renewable energy resources
and the analysis of climate anomalies, for example. However, for the
\nomenclature{GOC}{Golf of California} Gulf of California (GOC) to
date no single, reliable source of data exits to provide guidance
for such studies. Therefore, in this work the data from seven data
sources was combined to generate seasonal climatological means. These
are then provided in two formats: A two dimensional gridded dataset
of the temporally averaged data and temporally-spatially averaged
values.

\subsection{Geographic domains}

\begin{figure}
\begin{centering}
\subfloat[]{\protect\centering{}\protect\includegraphics[scale=0.5]{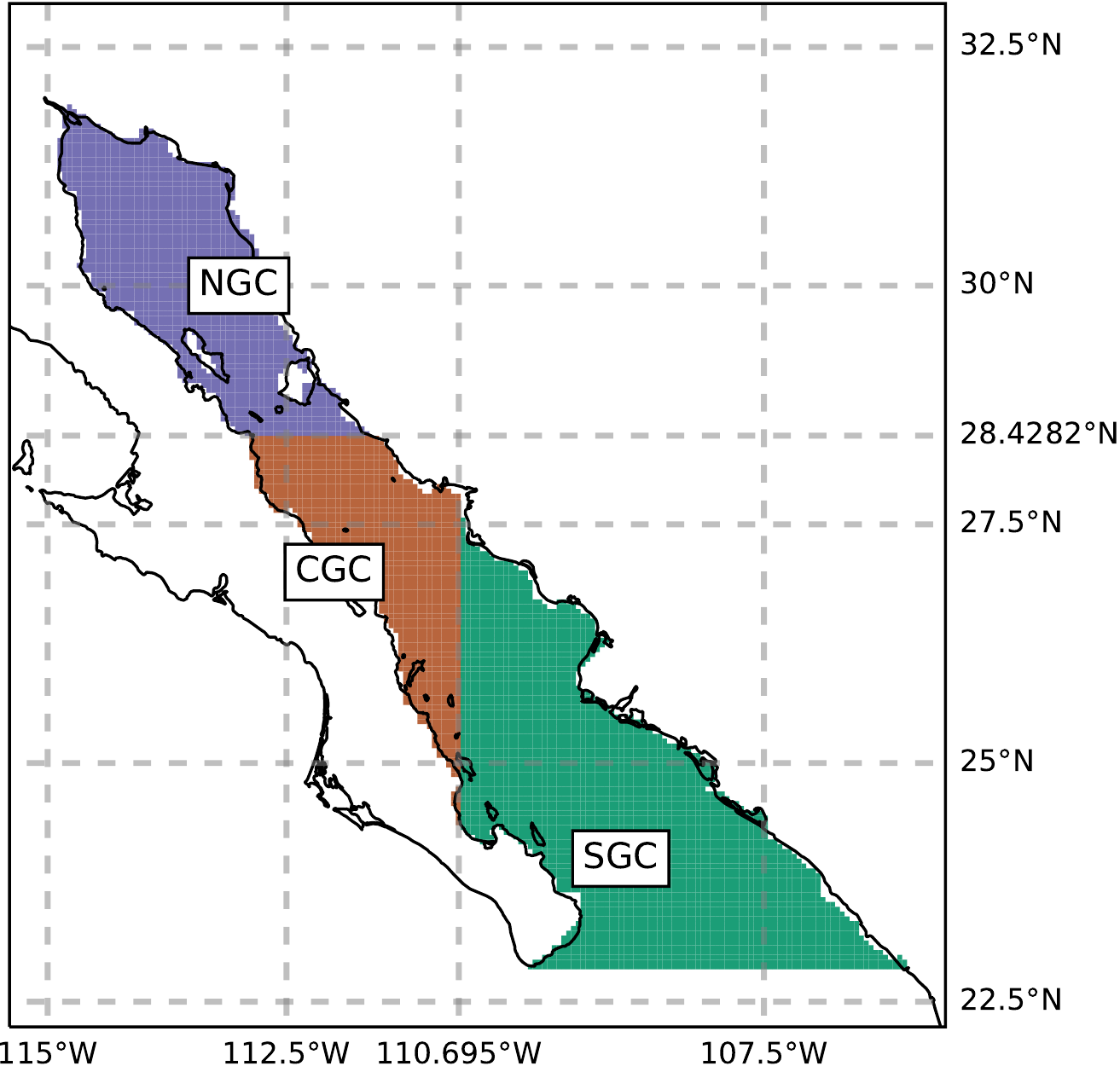}\protect}\quad{}\subfloat[]{\protect\centering{}\protect\includegraphics[scale=0.5]{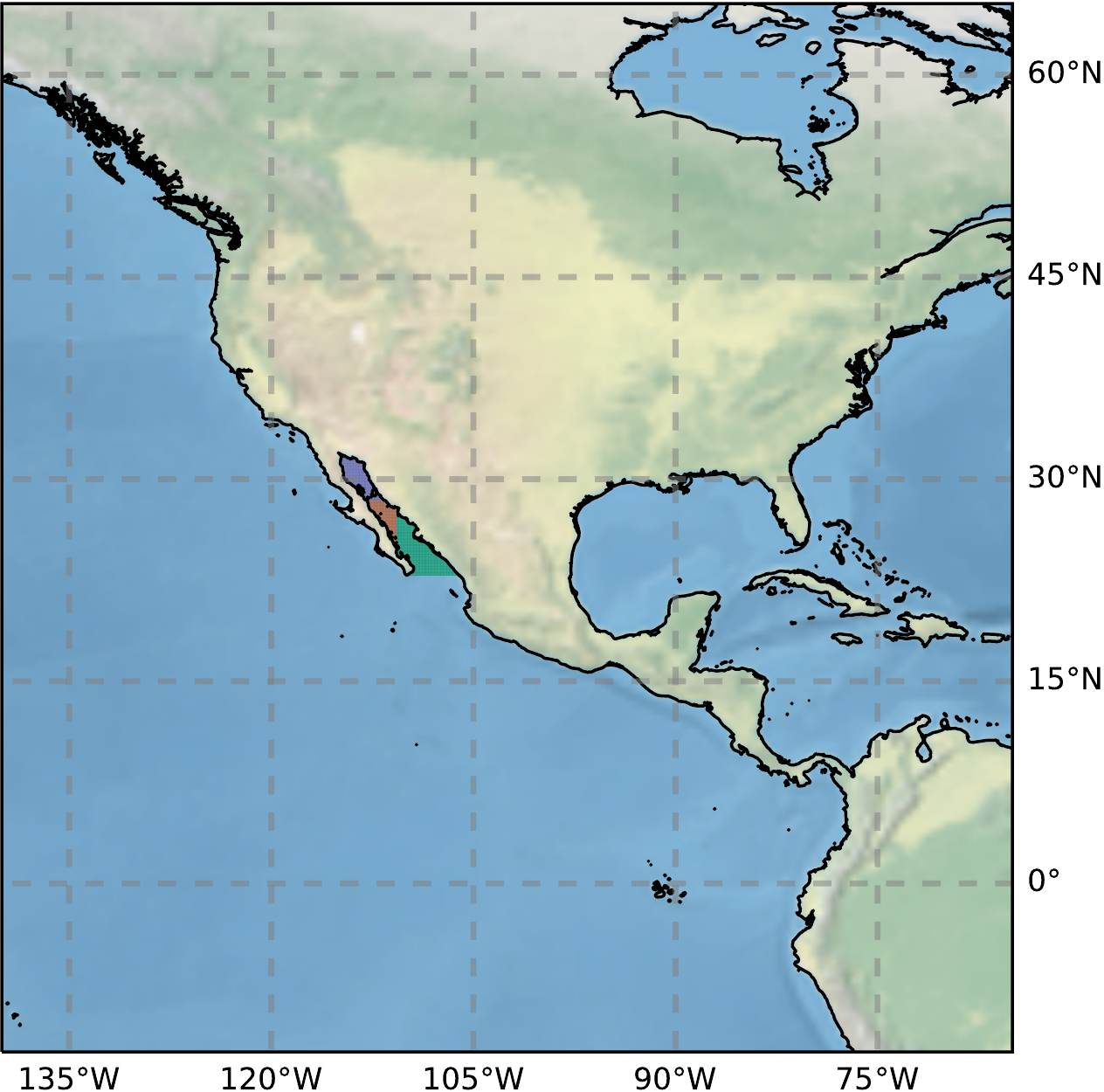}\protect}\protect\caption{Geographic domains for spatial averages \label{fig:Geographic-domains-for}}

\par\end{centering}

\end{figure}

The geographic domain is limited in the west by the Baja California
peninsular and in the east by the Mexican mainland. In the south the
data is considered until the last land point in the Baja California
peninsular. This domain is then further subdivided (sub-domains) into
the \nomenclature{NGC}{Northern Golf of California}Northern (NGC),
\nomenclature{CGC}{Central Golf of California}Central (CGC) and \nomenclature{SGC}{Southern Golf of California}Southern
Gulf (SGC), as shown in Figure \ref{fig:Geographic-domains-for}.
The southern limit to the NGC domain is latitude $28.428207\degree$.
CGC and SGC are divided by longitude $-110.695\degree$.

\section{Data production methods}

In the following the individual source dataset are described. Then
the processing steps to obtain the temporal and spatial mean dataset
are presented.

\subsection{Source datasets}

\begin{figure}

\begin{centering}
\includegraphics[scale=0.4]{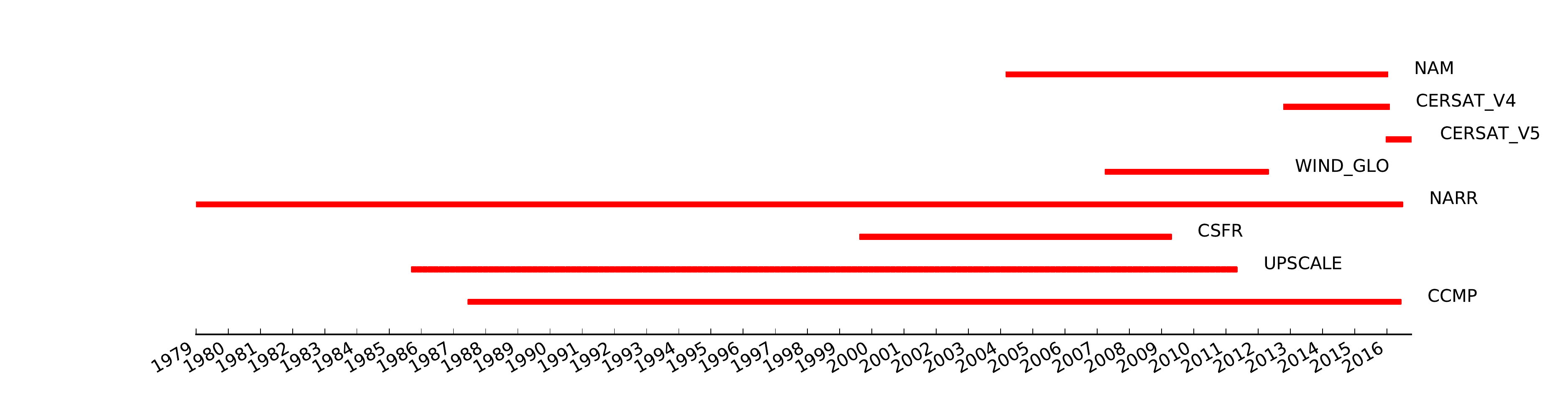}\protect\caption{Data source time coverage\label{fig:Data-sources}}

\par\end{centering}

\end{figure}

Four numerical models and three observational products were combined
in this study. The observational products were:
\begin{enumerate}
\item The Cross-Calibrated Multi-Platform observational product \nomenclature{CCMP}{Cross-Calibrated Multi-Platform}(CCMP), 
\item Global Wind, L4 2007-2012 Climatology (WIND\_GLO), 
\item Global Ocean - Wind Analysis - Blended Advanced Scatterometer (ASCAT)
- Special Sensor Microwave Imager (SSM/I) \nomenclature{SSM/I}{Special Sensor Microwave Imager}\nomenclature{ASCAT}{Advanced Scatterometer}
(CERSAT), 
\end{enumerate}
and the numerical models used were
\begin{enumerate}
\item UK Met Office Unified Model 8 UK on PRACE (the Partnership for Advanced
Computing in Europe) - weather-resolving Simulations of Climate for
global Environmental risk (UPSCALE),
\item Climate forecast system reanalysis (CFSR),
\item National Centers for Environmental Prediction (NCEP) North American
Regional Reanalysis (NARR) and the
\item North American Mesoscale model (NAM).
\end{enumerate}
The time coverage of the respective datasets is shown in Figure \ref{fig:Data-sources}
and details about each dataset are provided below.

\subsubsection{UPSCALE}

The UPSCALE dataset is the result of a global N512 (512 is the number
of 2 grid length waves that can be supported in the Eeast-West direction,
hence N512 yields 1024x768 grid points East-West x North-South) UPSCALE
project. It is based on the \nomenclature{HadGEM3}{Hadley Centre Global Environment Model version 3}\nomenclature{GA3}{Global Atmosphere 3}
Hadley Centre Global Environment Model version 3 (HadGEM3) - Global
Atmosphere (GA) 3.0 configuration of the Met Office Unified Model
\nomenclature{MetUM}{Met Office Unified Model}(MetUM) version 8.0,
combined with the \nomenclature{GL3.0}{Global Land 3.0} Global Land
(GL) 3.0 configuration of the \nomenclature{JULES}{Joint UK Land Environment Simulator}
Joint UK Land Environment Simulator (JULES) community land surface
model, as documented in \citet{gmd-4-919-2011}. The model was configured
to reproduce the present climate for the period Feb 1985 to Dec 2011.
More detailed information about the UPSCALE model simulations is available
in \citet{gmd-7-1629-2014,Donlon2012140}. Instantaneous winds, obtained
every three hours, are averaged over the respective seasons. The resolution
of the model was $0.23438\degree$ in latitude and $0.3516\degree$
degrees in longitude.

\subsubsection{NAM-ANL}

The \nomenclature{NAM}{North American Mesoscale Forecast System}
North American Mesoscale Forecast System (NAM) dataset is a \nomenclature{NWP}{Numerical Weather Prediction}numerical
weather prediction (NWP) dataset produced by the US National Weather
Service - \nomenclature{NCEP}{National Centers for Environmental Prediction}
National Centers for Environmental Prediction (NCEP), for North America,
using the MESO ETA Model, \citet{doi:10.1175/1520-0434(1994)009<0265:TNNMEM>2.0.CO;2}.
The grid resolution is 12km using the Advanced Weather Interactive
Processing System (AWIPS) lambert conformal grid over the Contiguous
United States (CONUS). The raw data was re-gridded (using bi-linear
interpolation) onto a lat-long grid with uniform $0.05\degree$ grid
spacing. Instantaneous winds, obtained every six hours, are averaged
over the respective seasons.

\subsubsection{NARR}

NCEP \nomenclature{NARR}{North American Regional Analysis}North American
Regional Analysis (NARR) dataset, \citet{doi:10.1175/BAMS-87-3-343},
is a regional reanalysis dataset. The monthly means of the original
dataset are averaged over the respective seasons. 450 monthly mean
winds at $10$m were used. The NARR model uses the NCEP Eta Model
(32km Lambert Conformal/45 layer) combined with the \nomenclature{RDAS}{Regional Data Assimilation System}Regional
Data Assimilation System (RDAS). The resulting dataset improves significantly
on the accuracy of temperature, winds and precipitation compared to
the NCEP-DOE Global Reanalysis 2, \citet{doi:10.1175/BAMS-83-11-1631}.
Here the data was re-gridded onto a $0.2\degree$ uniform lat-long
grid.

\subsubsection{CCMP}

The Cross-Calibrated Multi-Platform \nomenclature{CCMP}{Cross-Calibrated Multi-Platforr}(CCMP)
ocean surface wind vector analyses (\citet{doi:10.1175/2010BAMS2946.1,doi:10.1175/1520-0477(1996)077<0869:AMGSWV>2.0.CO;2})
provide a consistent, gap-free long-term time-series from July 1987
through June 2011. The CCMP datasets combine cross-calibrated satellite
winds using a Variational Analysis Method (VAM), \citet{doi:10.1175/1520-0426(2003)20<585:ATDVAM>2.0.CO;2},
to produce a $0.25\degree$ gridded analysis. The CCMP dataset uses
satellite winds derived by Remote Sensing Systems (RSS) from a number
of microwave satellite instruments. RSS applies a sea-surface emissivity
model and radiative transfer function to derive surface winds. Wind
speeds and directions from microwave scatterometers (including \nomenclature{QuikScat}{NASA's Quik Scatterometer }NASA's
Quik Scatterometer (QuikScat) and its SeaWinds instrument) are also
considered. Both radiometer and scatterometer data are validated against
ocean moored buoys. The VAM combines the RSS data with in situ measurements
and a starting estimate of the wind field. The European Center for
Medium-Range Weather Forecasts \nomenclature{ECMWF}{European Center for Medium-Range Weather Forecasts}(ECMWF)
ERA-40 Reanalysis is used as the first-guess from 1987 to 1998. The
ECMWF Operational analysis is used from January 1999 onward. All wind
observations and analysis fields are referenced to a height of $10$
meters. The landmask for this dataset was generated from the topography
of the UPSCALE dataset, re-gridded to the CCMP resolution.

\subsubsection{CSFR- CLIMATE FORECAST SYSTEM REANALYSIS}

The CFSR, \citet{doi:10.1175/2010BAMS3001.1}, includes coupling of
atmosphere and ocean during the generation of the 6 hour guess field,
an interactive sea-ice model, and assimilation of satellite radiances.
The CFSR global atmosphere resolution is $\approx38$ km (T382) with
$64$ vertical levels. The global ocean is $0.25\degree$ at the equator,
extending to a global $0.5\degree$ beyond the tropics, with $40$
vertical levels. Ocean-atmosphere interactions are not used directly.
Rather the information is used for background information. The actual
reanalysis is uncoupled. $117$ monthly averages of the six-hourly
analyses were included in this study.

\subsubsection{CERSAT (V5 and V3)}

The CERSAT (Centre ERS d'Archivage et de Traitement) Global Blended
Mean Wind Fields V5 include wind components (meridional and zonal),
wind module and wind stress. They are estimated from scatterometers
ASCAT and Oceansat-2 Scatterometer (OSCAT) retrievals and from ECMWF
operational wind analysis with a horizontal resolution of $0.25\degree$
and $6$ hours in time. The estimation and calibration of the wind
product made use of ASCAT and OSCAT scatterometer swath winds, ECMWF
wind analysis, and moored buoy data. For the period 2012-2015 the
V3 dataset was used.

\subsubsection{GLO}

The Global Ocean CERSAT surface wind climatology, covering the years
2007-2012, is estimated from ASCAT retrievals \citet{doi:10.1080/01431161.2011.600348}.
The analyses are estimated as monthly averaged data with spatial resolution
of $0.25\degree$ in latitude and longitude.

\subsection{Processing of source data}

First the data was averaged in time. Individual netcdf files are provided
with the seasonal temporal averages, for each source dataset. Then
the spatial averages were computed. At first this averaging was performed
individually for each dataset. Following a selection procedure outlined
below the selected averages were then combined to form one single
representative value for each season and geographic sub-domain, respectively.

\subsubsection{Temporal-Spatial averages of the datasets}

\begin{figure}
\begin{centering}
\subfloat[]{\protect\begin{centering}
\protect\includegraphics[scale=0.25]{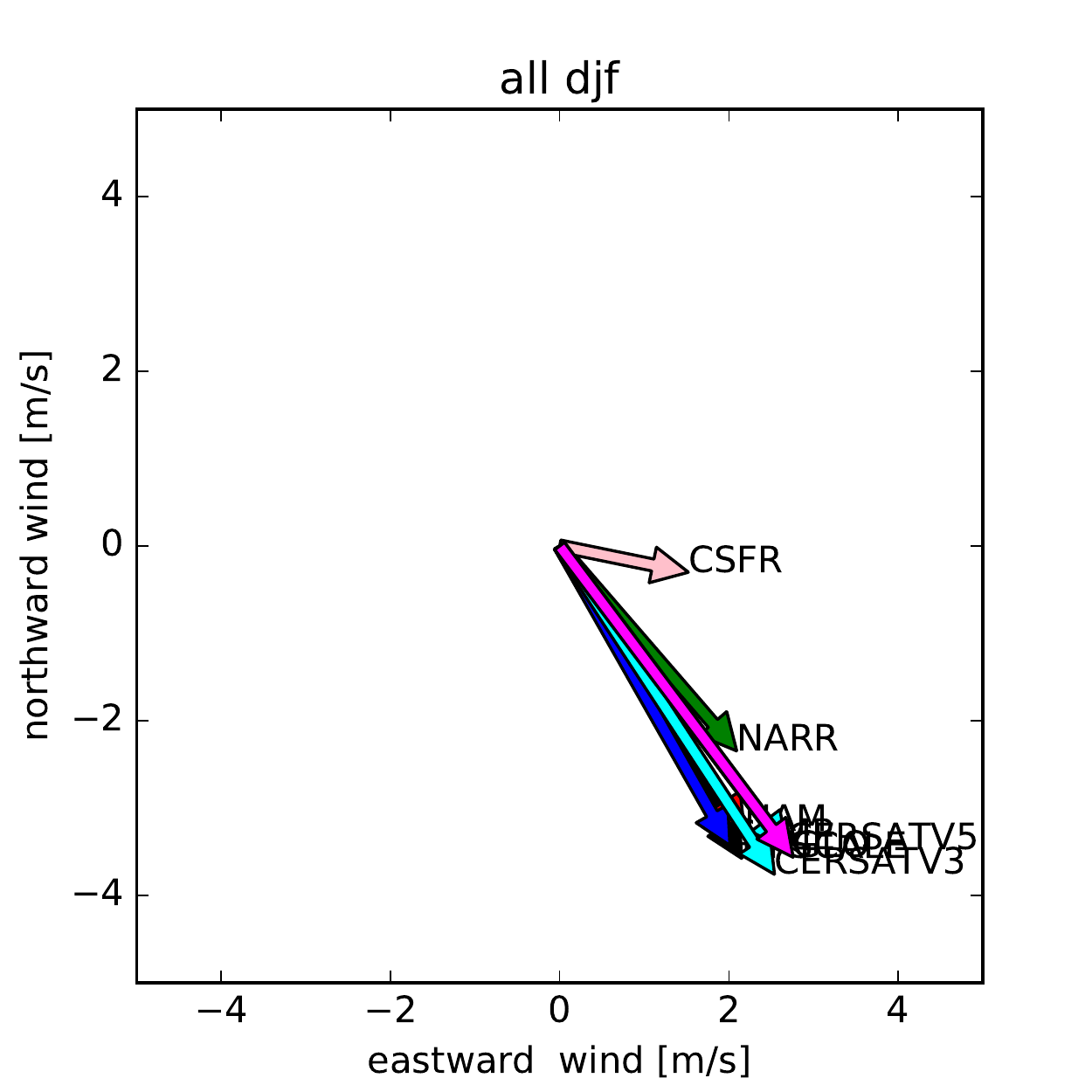}\protect
\par\end{centering}

}\subfloat[]{\protect\centering{}\protect\includegraphics[scale=0.25]{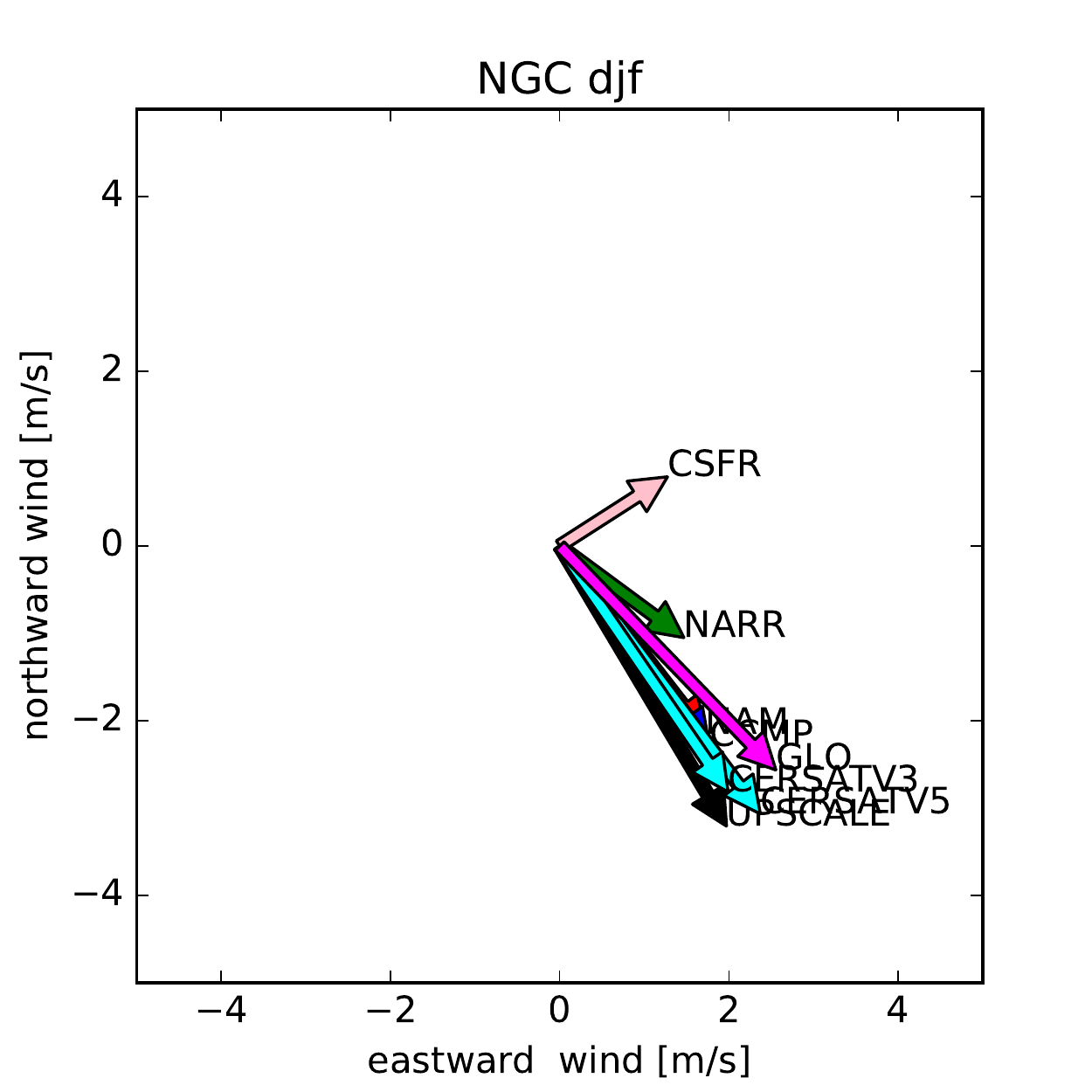}\protect}\subfloat[]{\protect\centering{}\protect\includegraphics[scale=0.25]{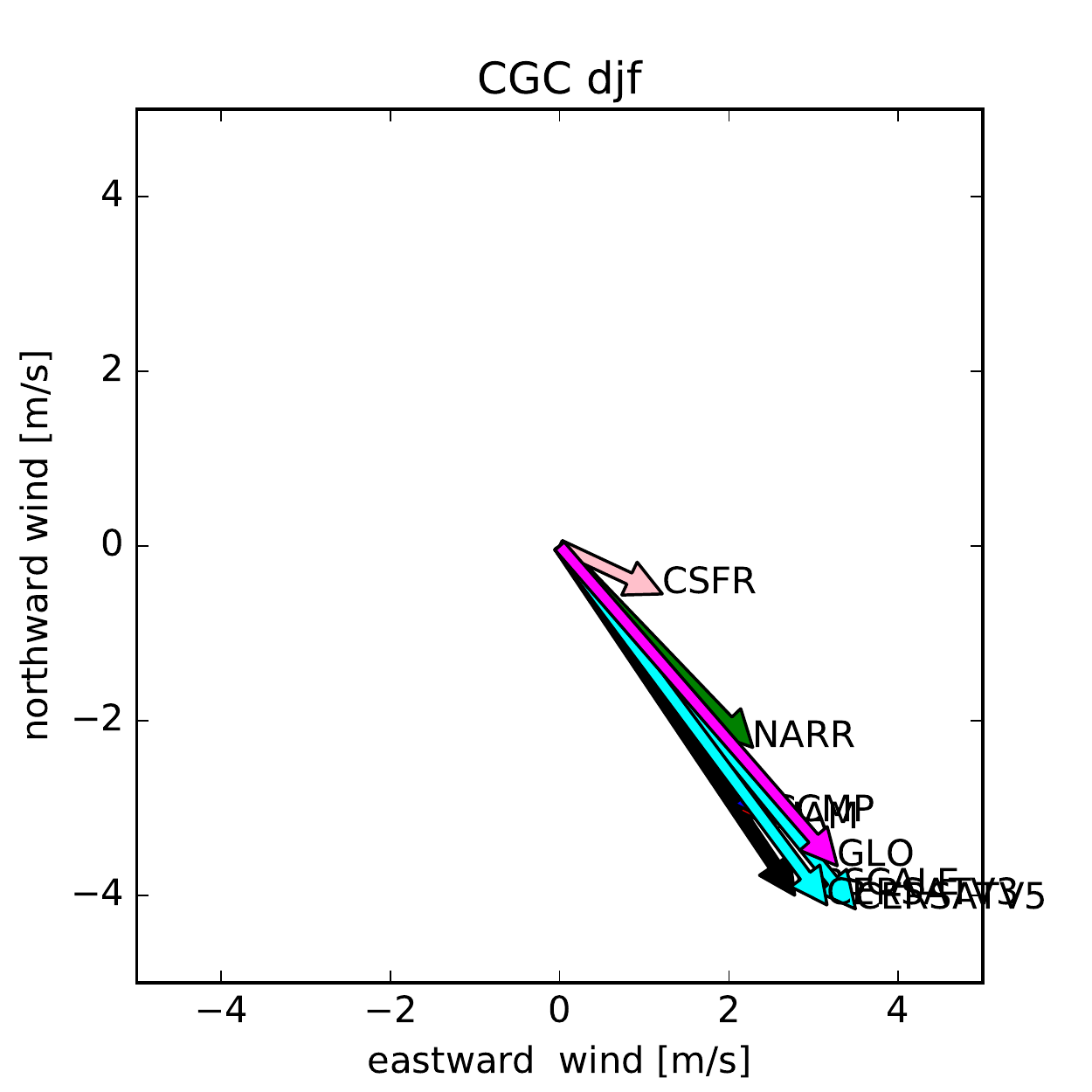}\protect}\subfloat[]{\protect\centering{}\protect\includegraphics[scale=0.25]{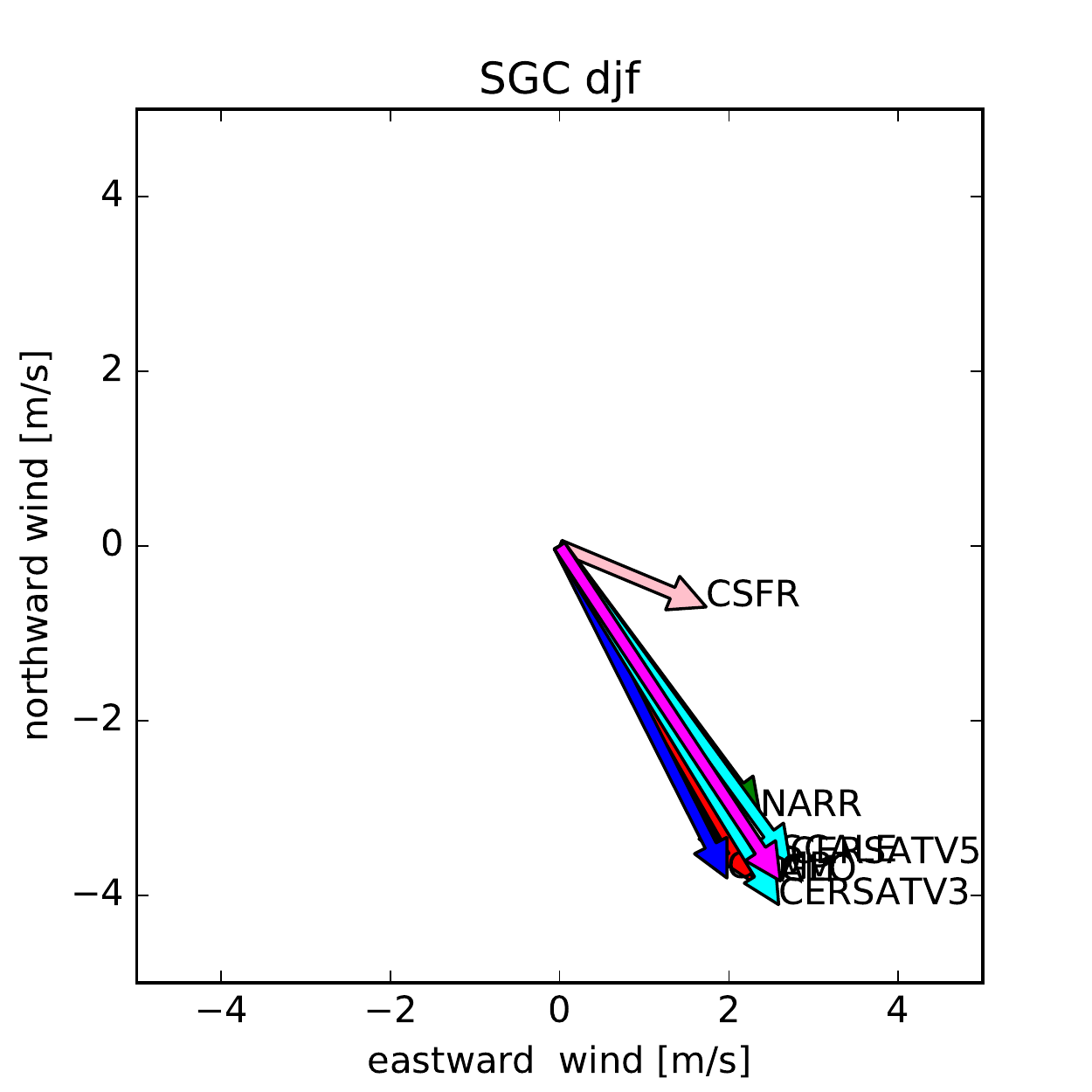}\protect}
\par\end{centering}

\begin{centering}
\subfloat[]{\protect\centering{}\protect\includegraphics[scale=0.25]{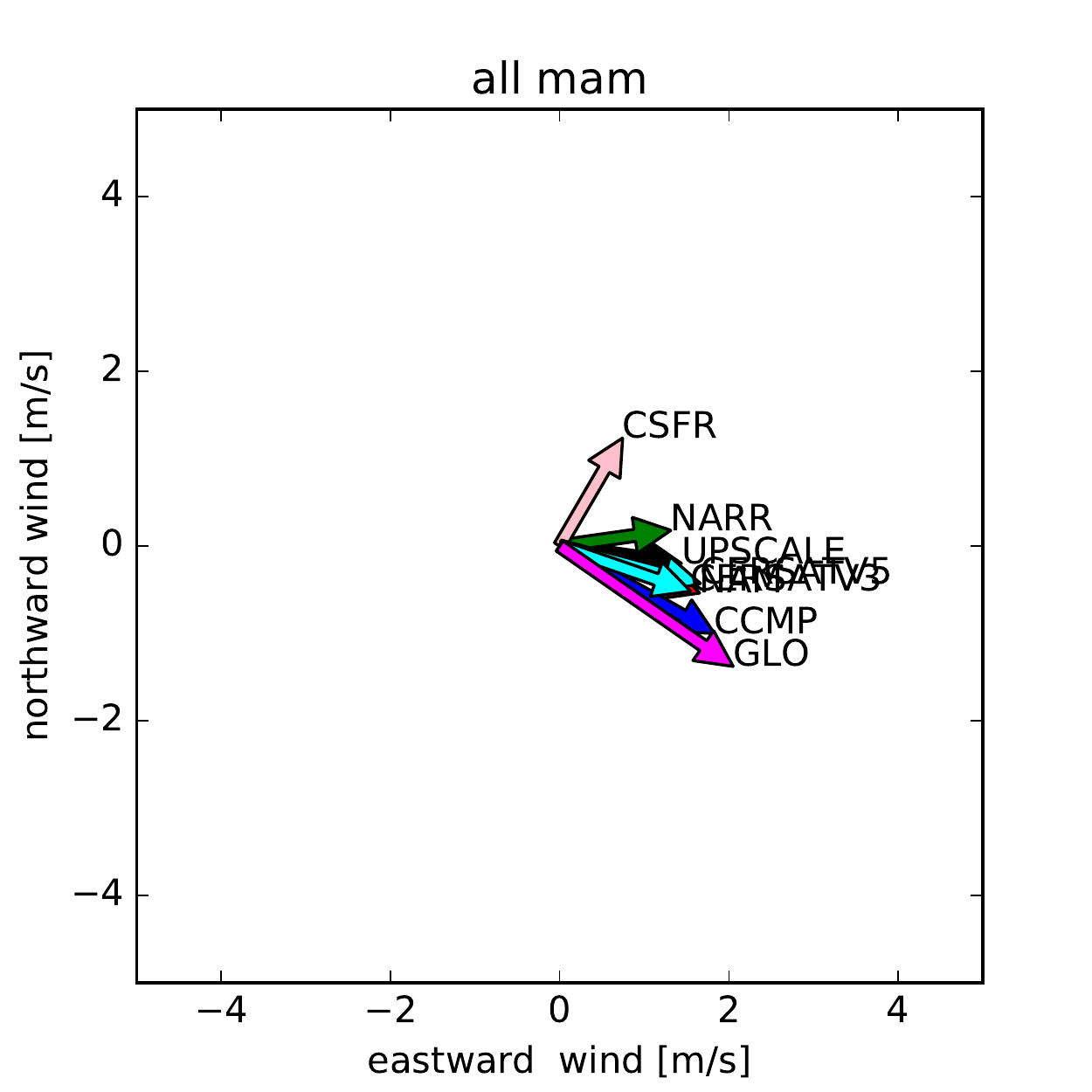}\protect}\subfloat[]{\protect\centering{}\protect\includegraphics[scale=0.25]{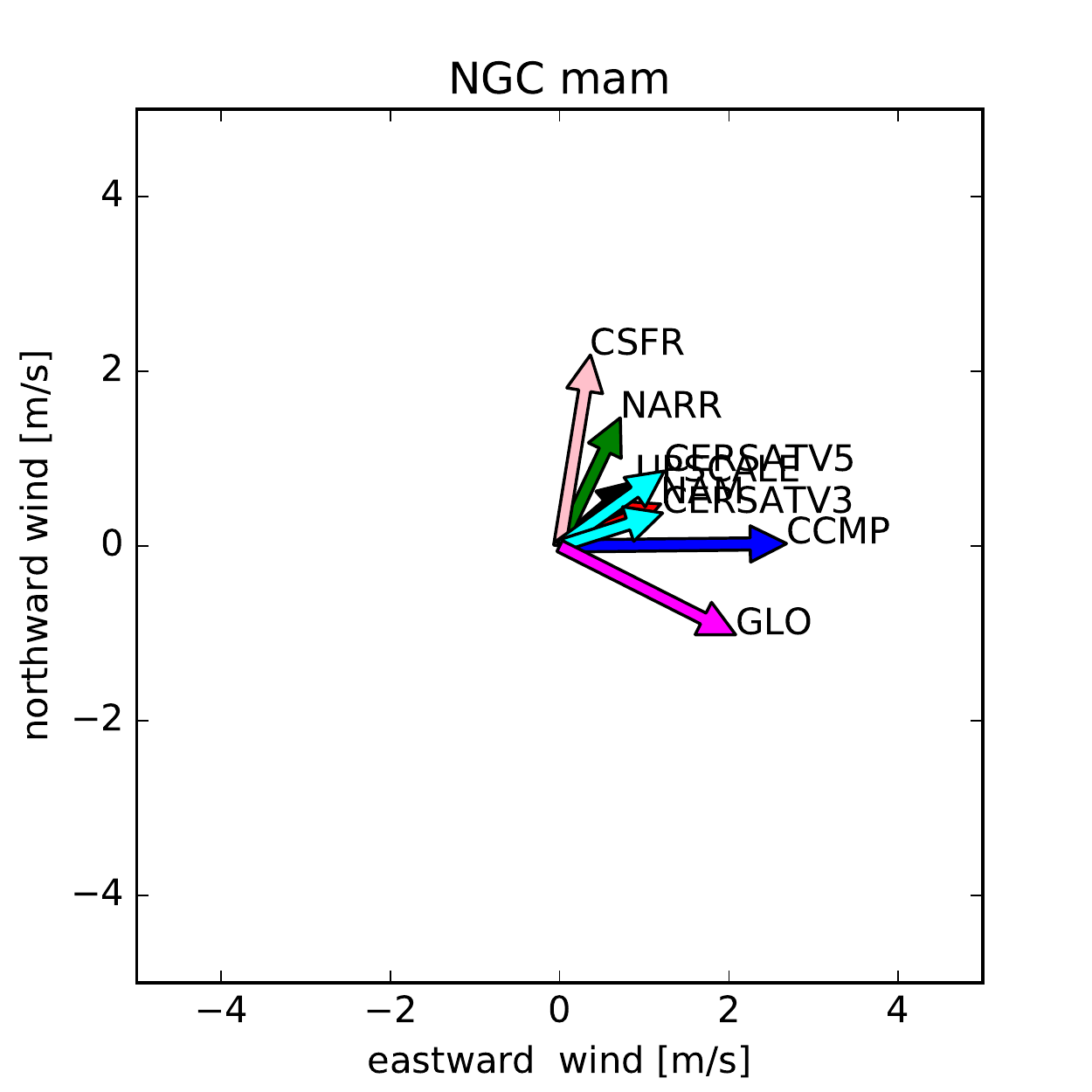}\protect}\subfloat[]{\protect\centering{}\protect\includegraphics[scale=0.25]{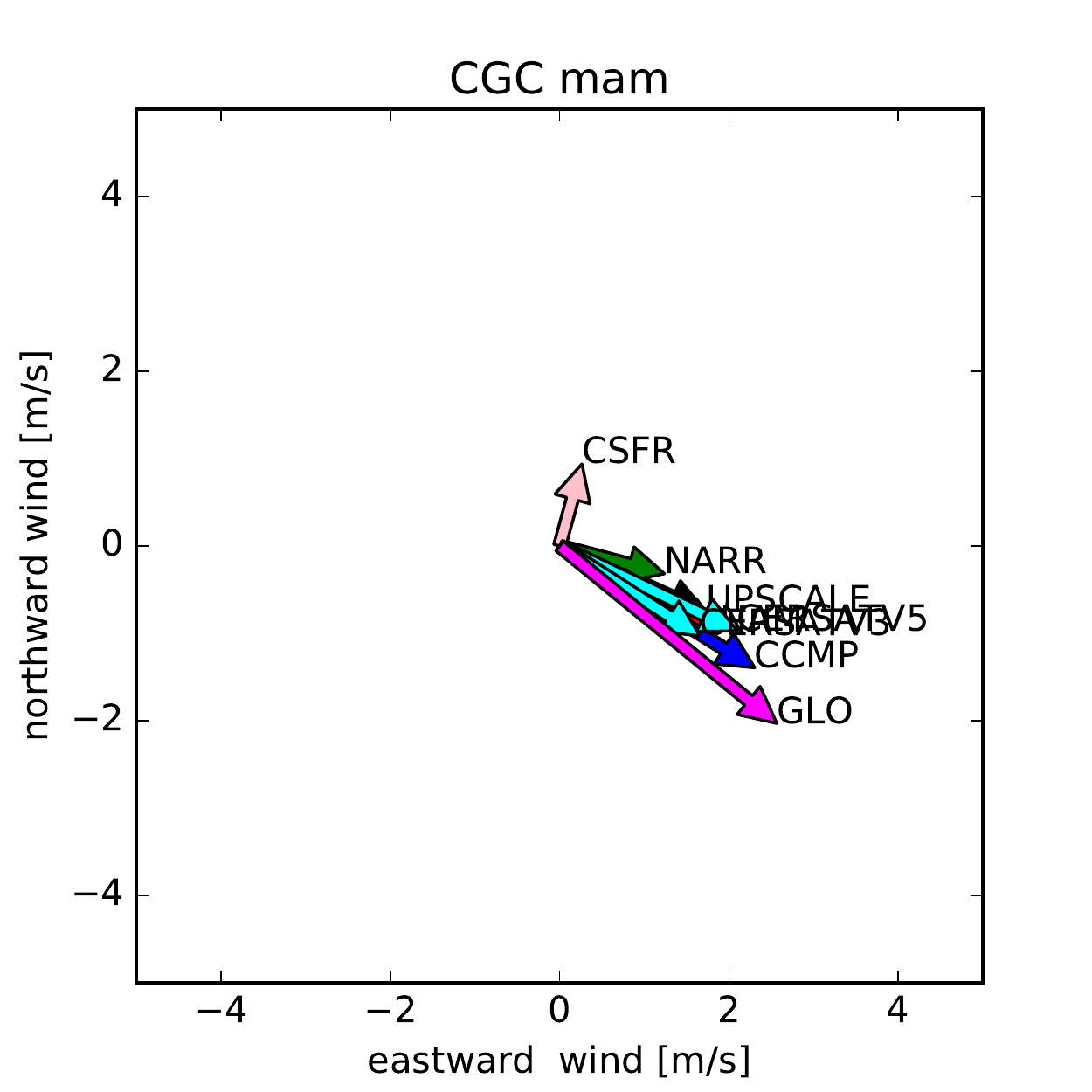}\protect}\subfloat[]{\protect\centering{}\protect\includegraphics[scale=0.25]{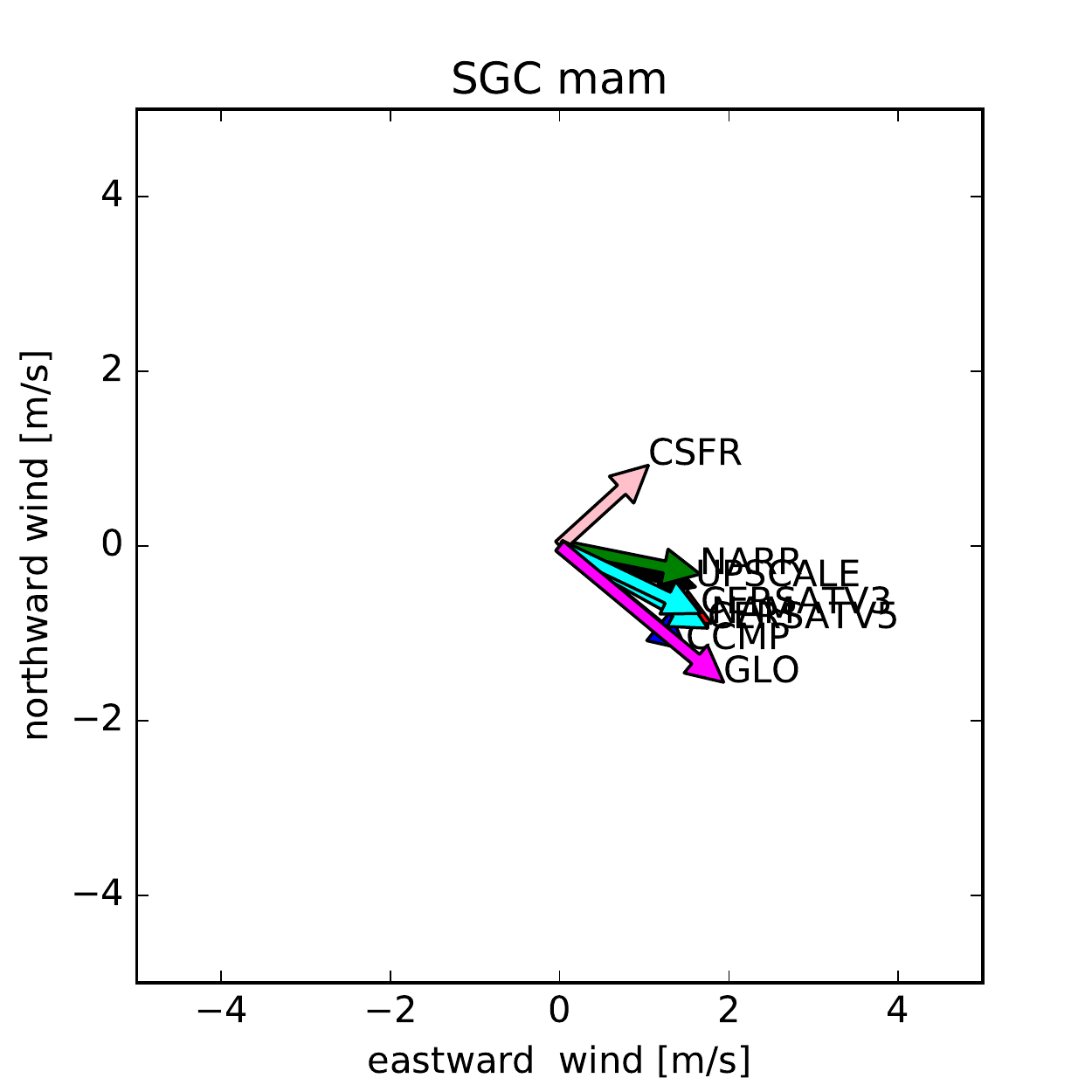}\protect}
\par\end{centering}

\begin{centering}
\subfloat[]{\protect\centering{}\protect\includegraphics[scale=0.25]{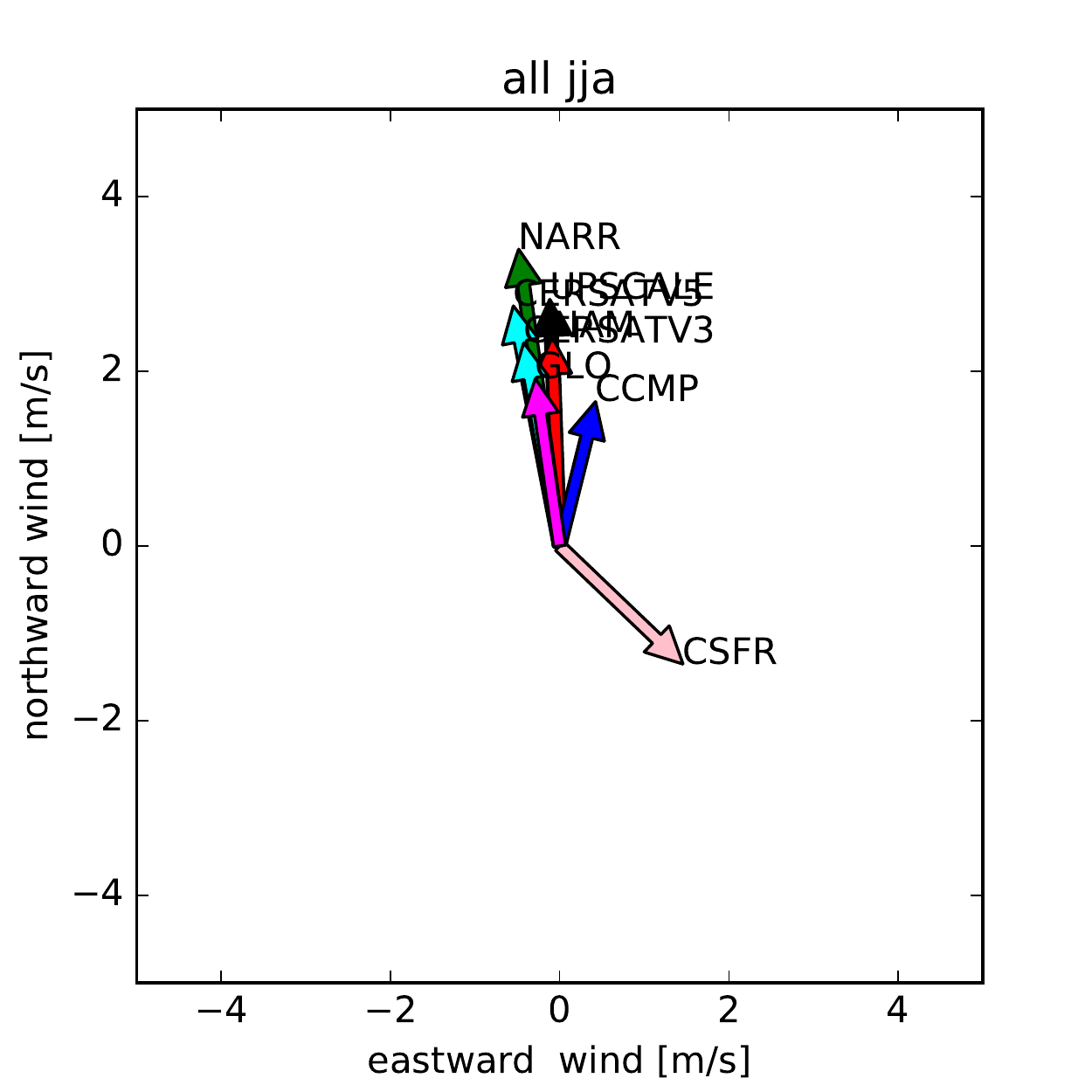}\protect}\subfloat[]{\protect\centering{}\protect\includegraphics[scale=0.25]{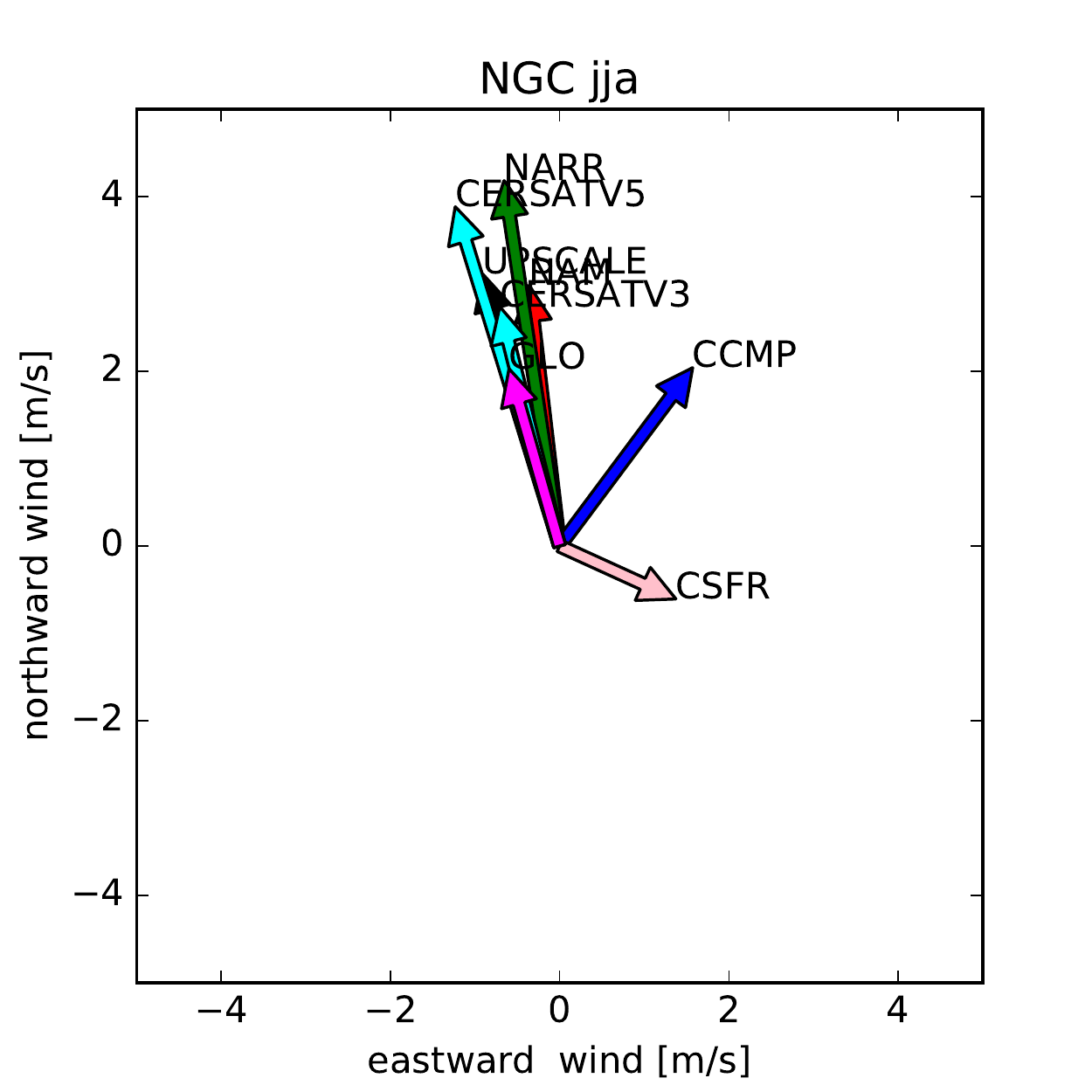}\protect}\subfloat[]{\protect\centering{}\protect\includegraphics[scale=0.25]{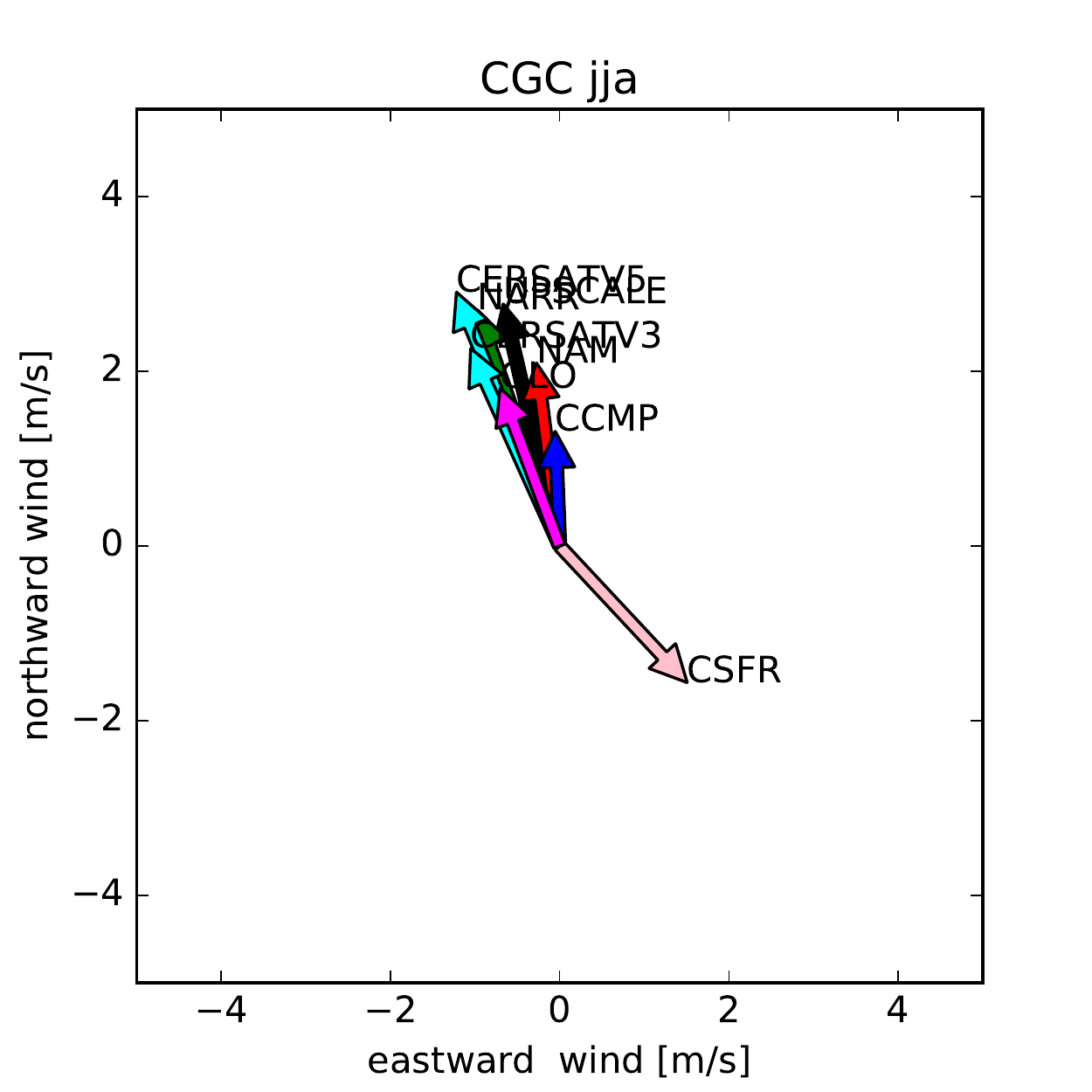}\protect}\subfloat[]{\protect\centering{}\protect\includegraphics[scale=0.25]{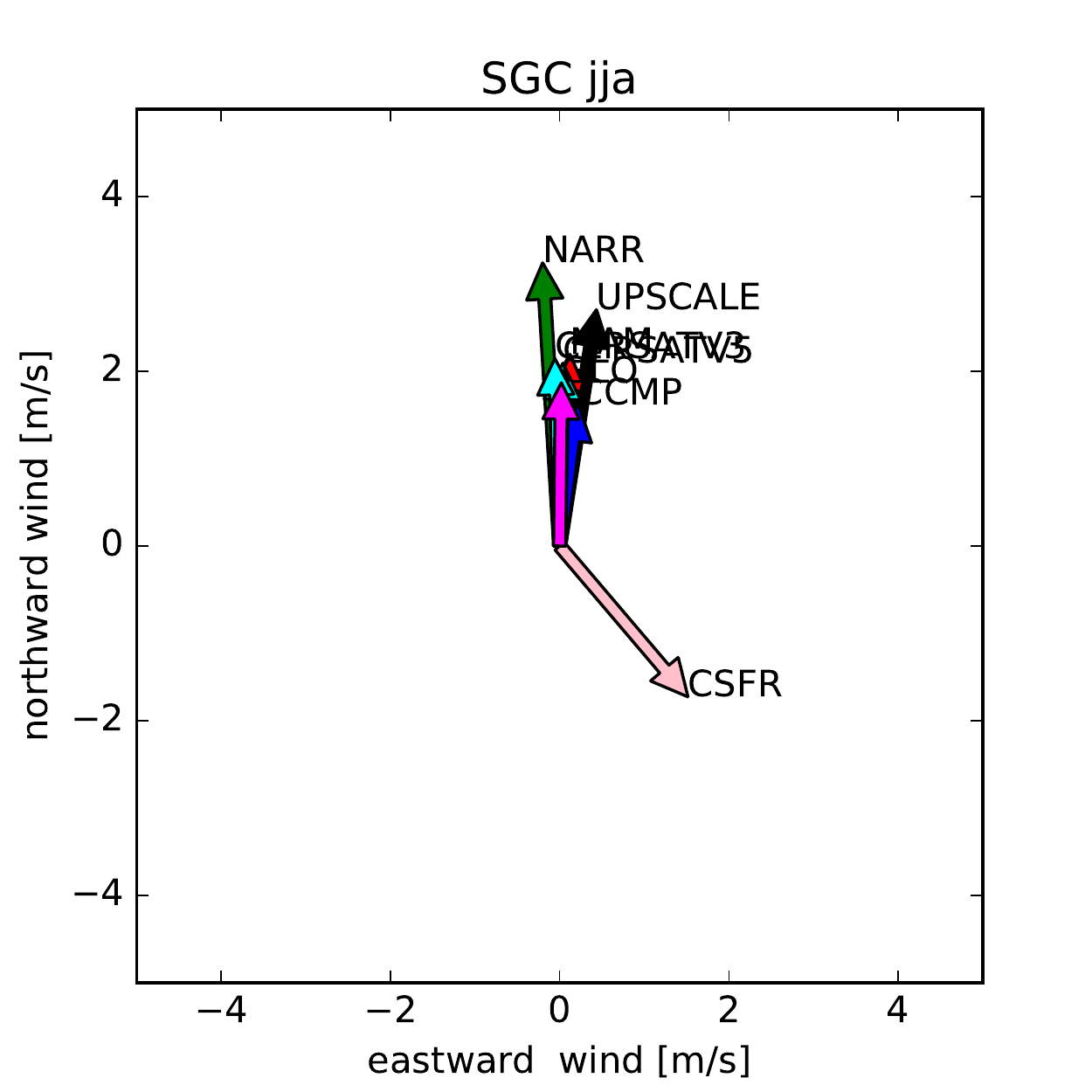}\protect}
\par\end{centering}

\begin{centering}
\subfloat[]{\protect\centering{}\protect\includegraphics[scale=0.25]{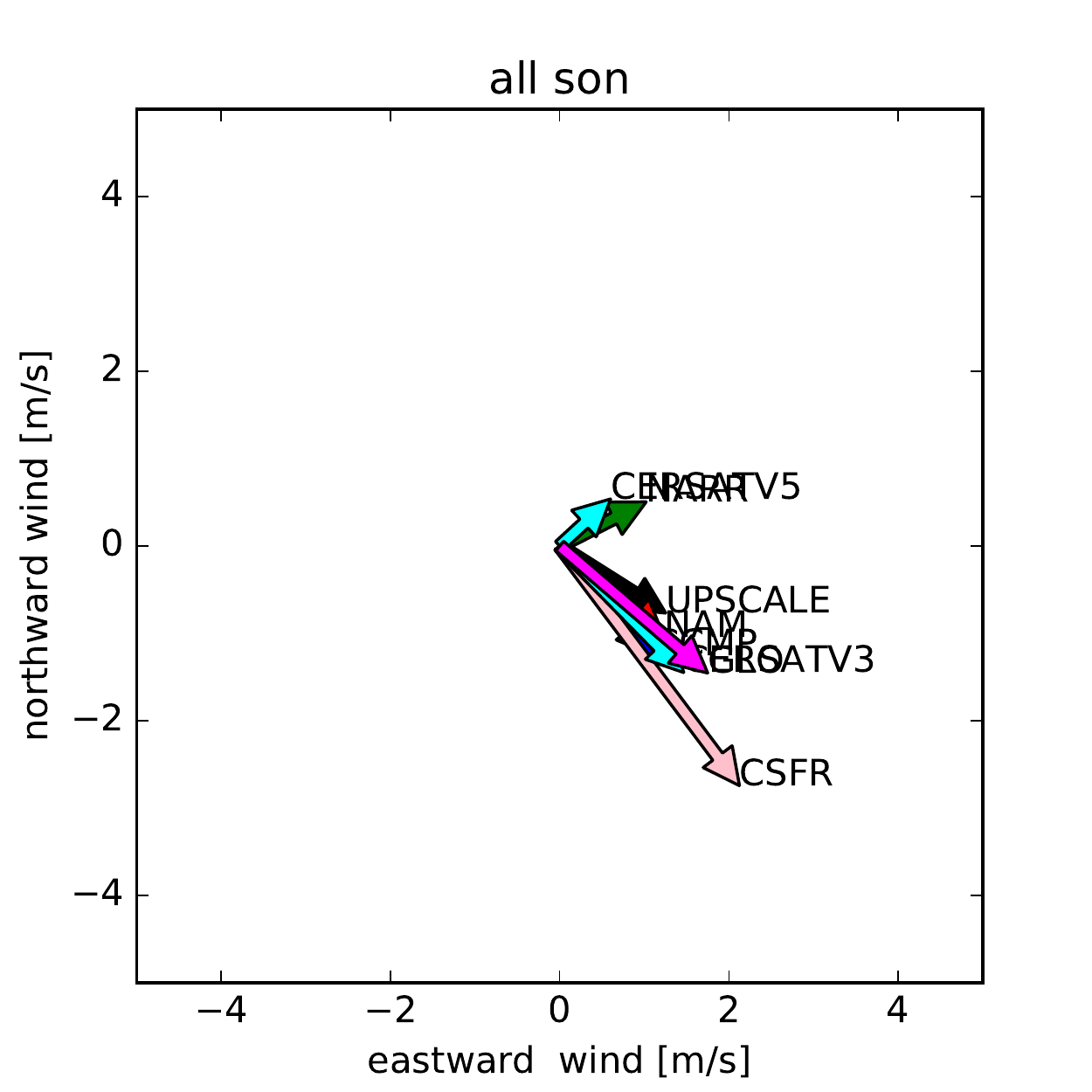}\protect}\subfloat[]{\protect\centering{}\protect\includegraphics[scale=0.25]{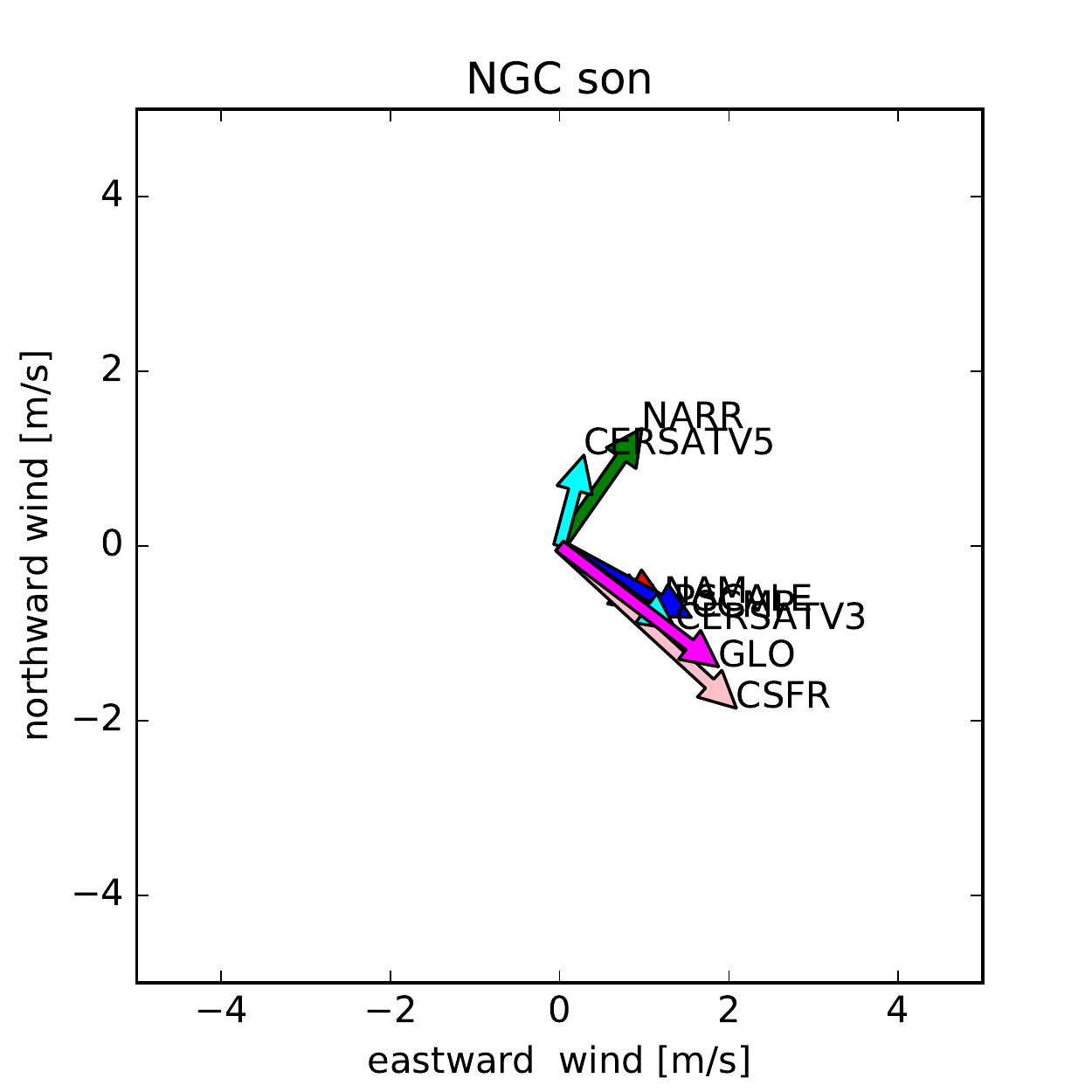}\protect}\subfloat[]{\protect\centering{}\protect\includegraphics[scale=0.25]{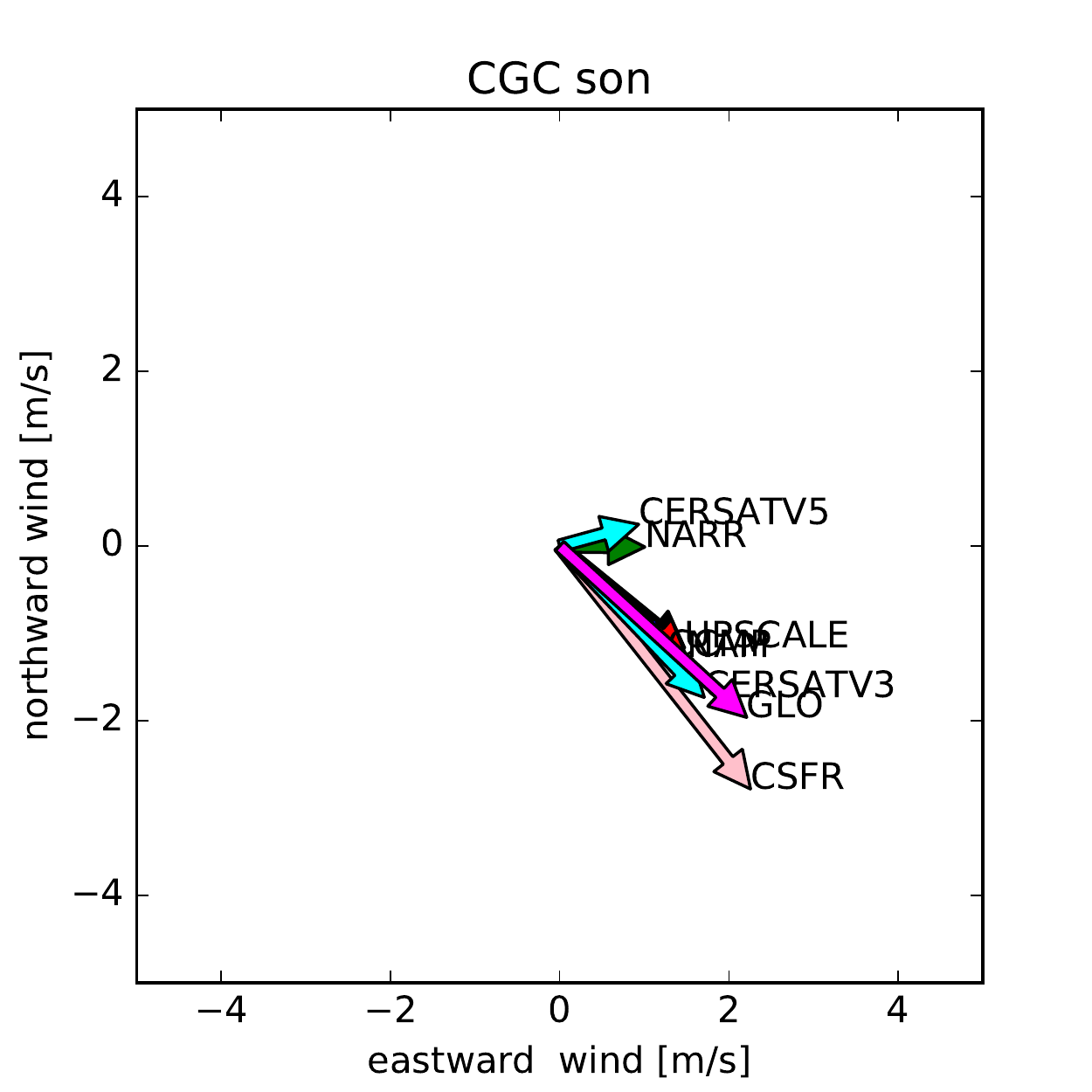}\protect}\subfloat[]{\protect\centering{}\protect\includegraphics[scale=0.25]{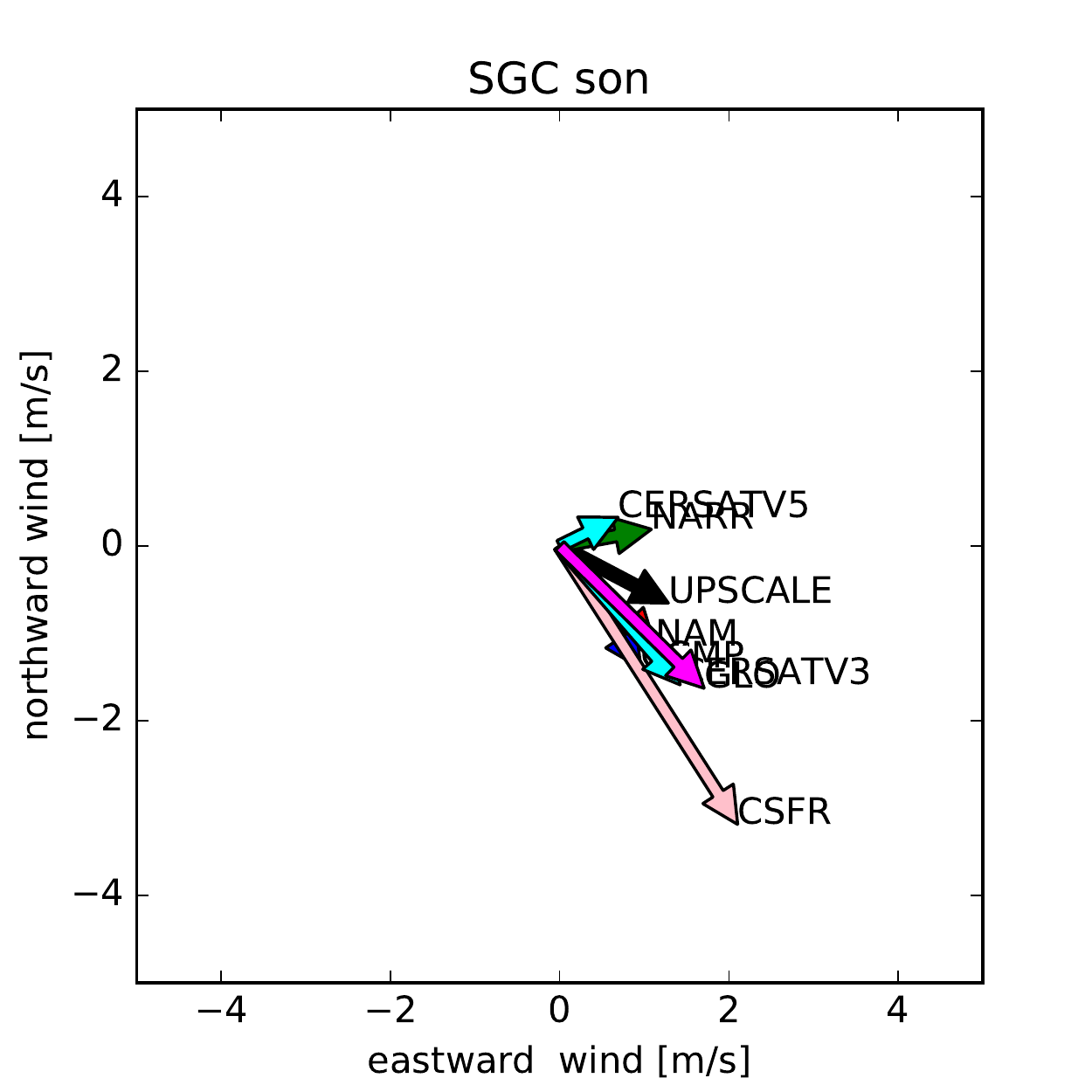}\protect}
\par\end{centering}

\centering{}\protect\caption{Temporal-spatial averages\label{fig:Temporal-spatial-averages}}
\end{figure}

The datasets were averaged in space according to the three regions
presented above. No particular weighting was applied as the difference
in areas covered north to south is small. The resulting average wind
speed and direction is illustrated in Figure \ref{fig:Temporal-spatial-averages}
for each sub-domain and season.

\subsubsection{Data selection}

\begin{figure}
\begin{centering}
\subfloat[]{\protect\centering{}\protect\includegraphics[scale=0.25]{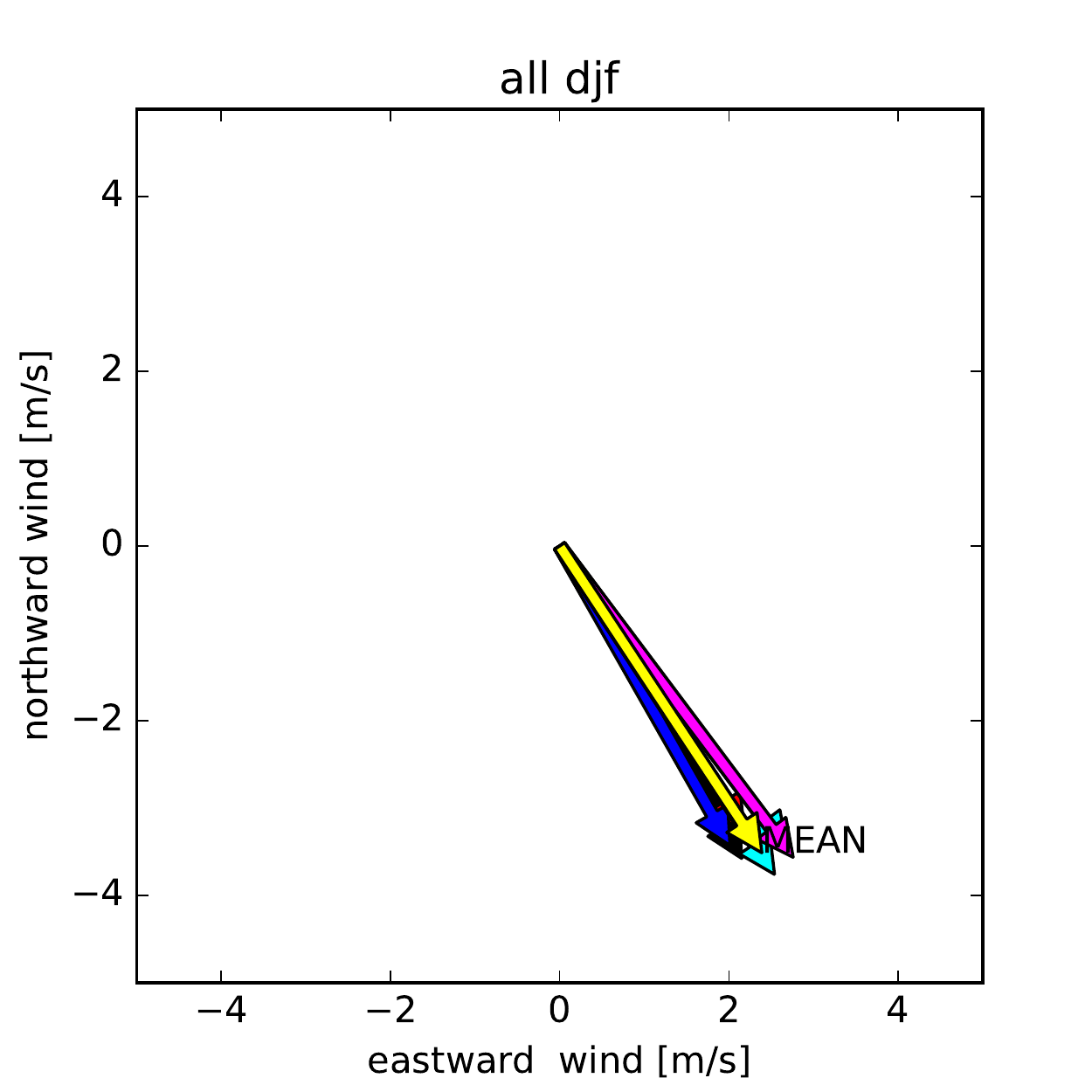}\protect}\subfloat[]{\protect\centering{}\protect\includegraphics[scale=0.25]{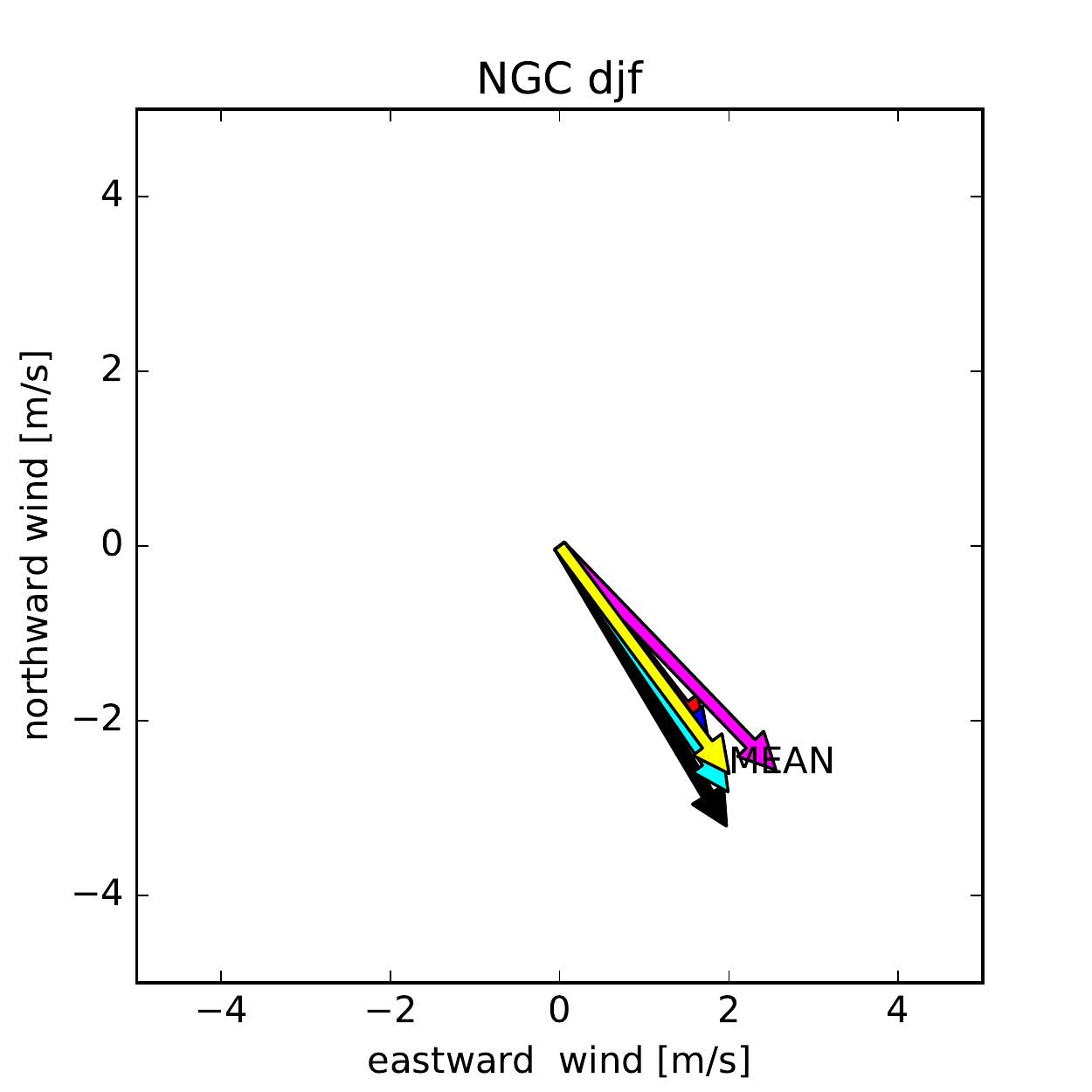}\protect}\subfloat[]{\protect\centering{}\protect\includegraphics[scale=0.25]{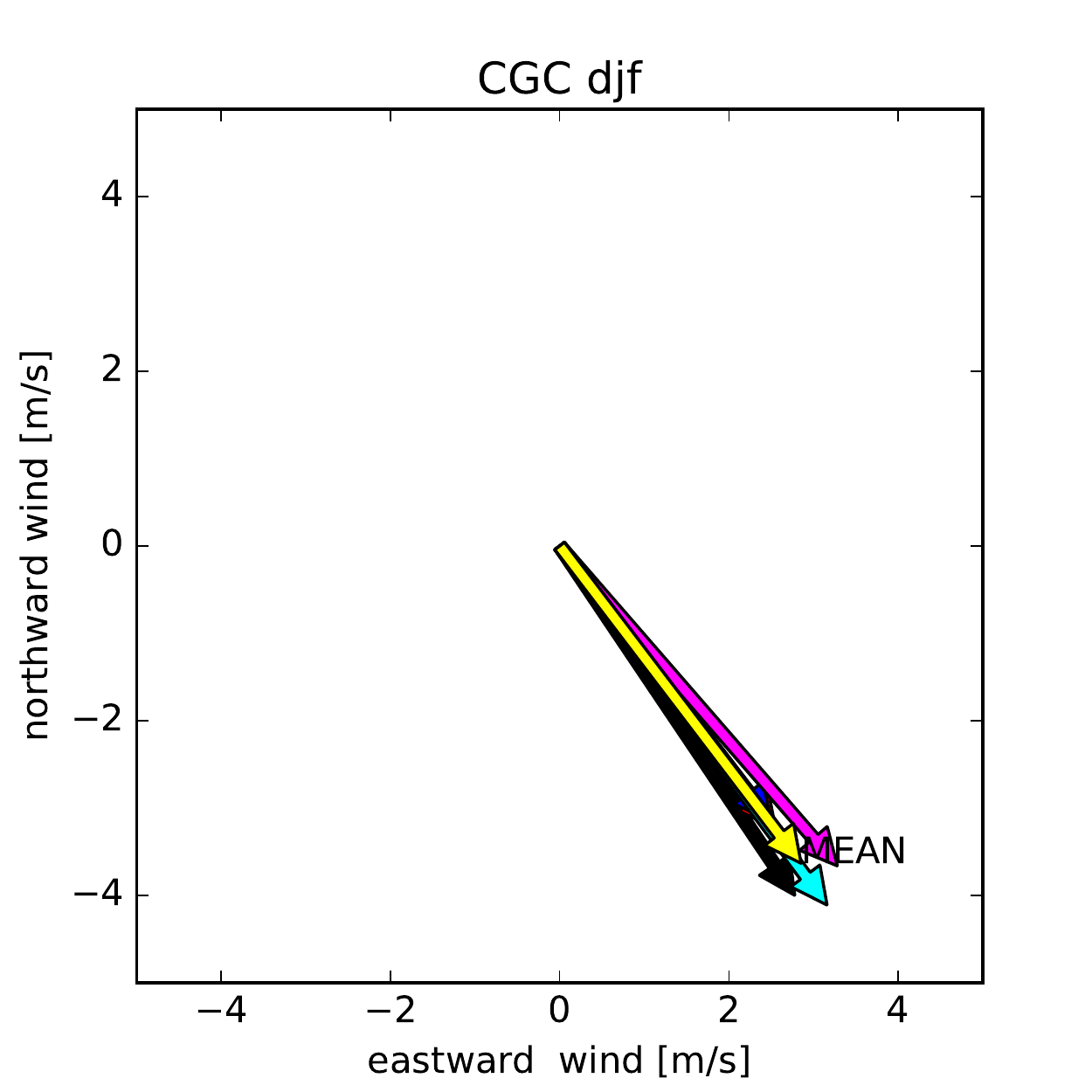}\protect}\subfloat[]{\protect\centering{}\protect\includegraphics[scale=0.25]{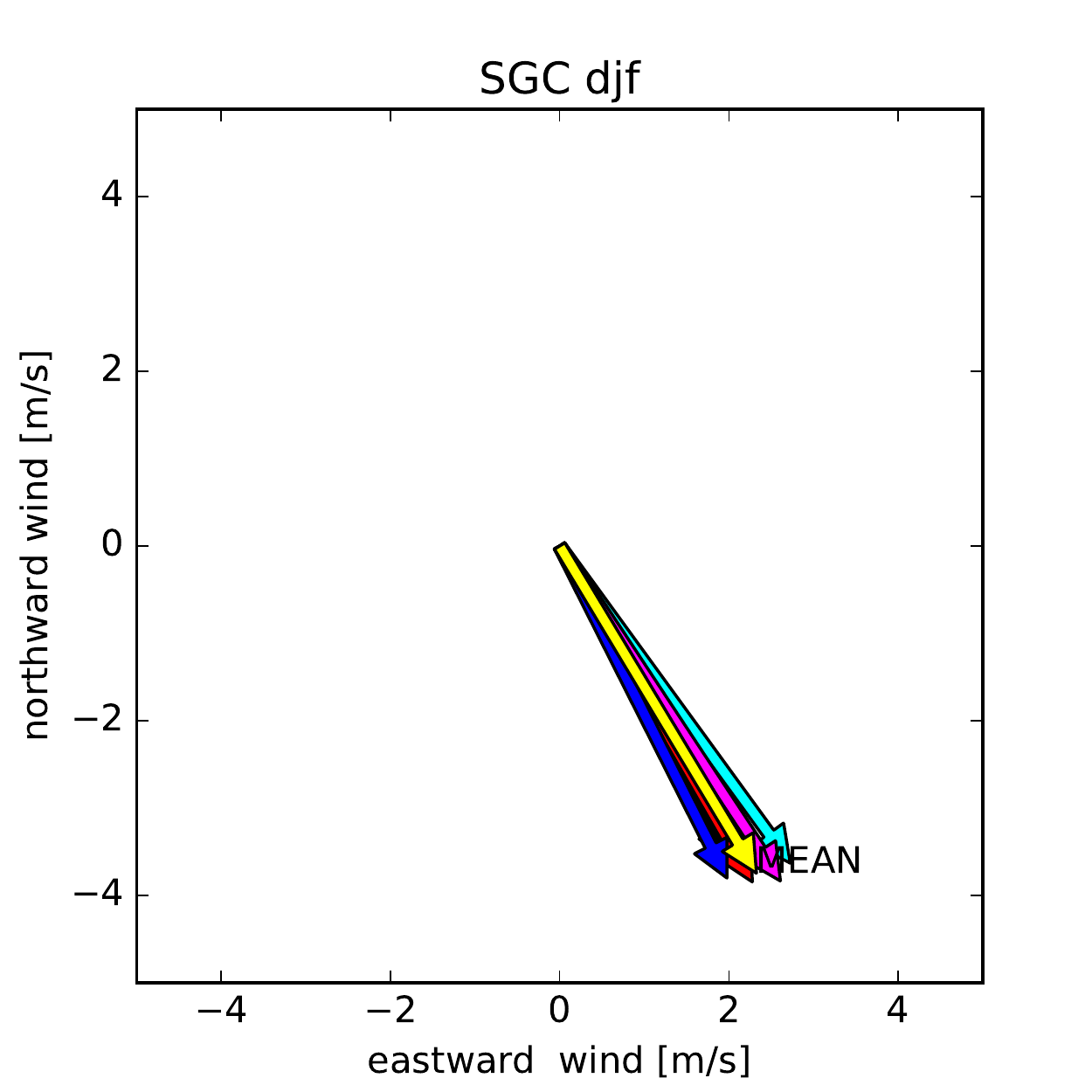}\protect}
\par\end{centering}

\begin{centering}
\subfloat[]{\protect\centering{}\protect\includegraphics[scale=0.25]{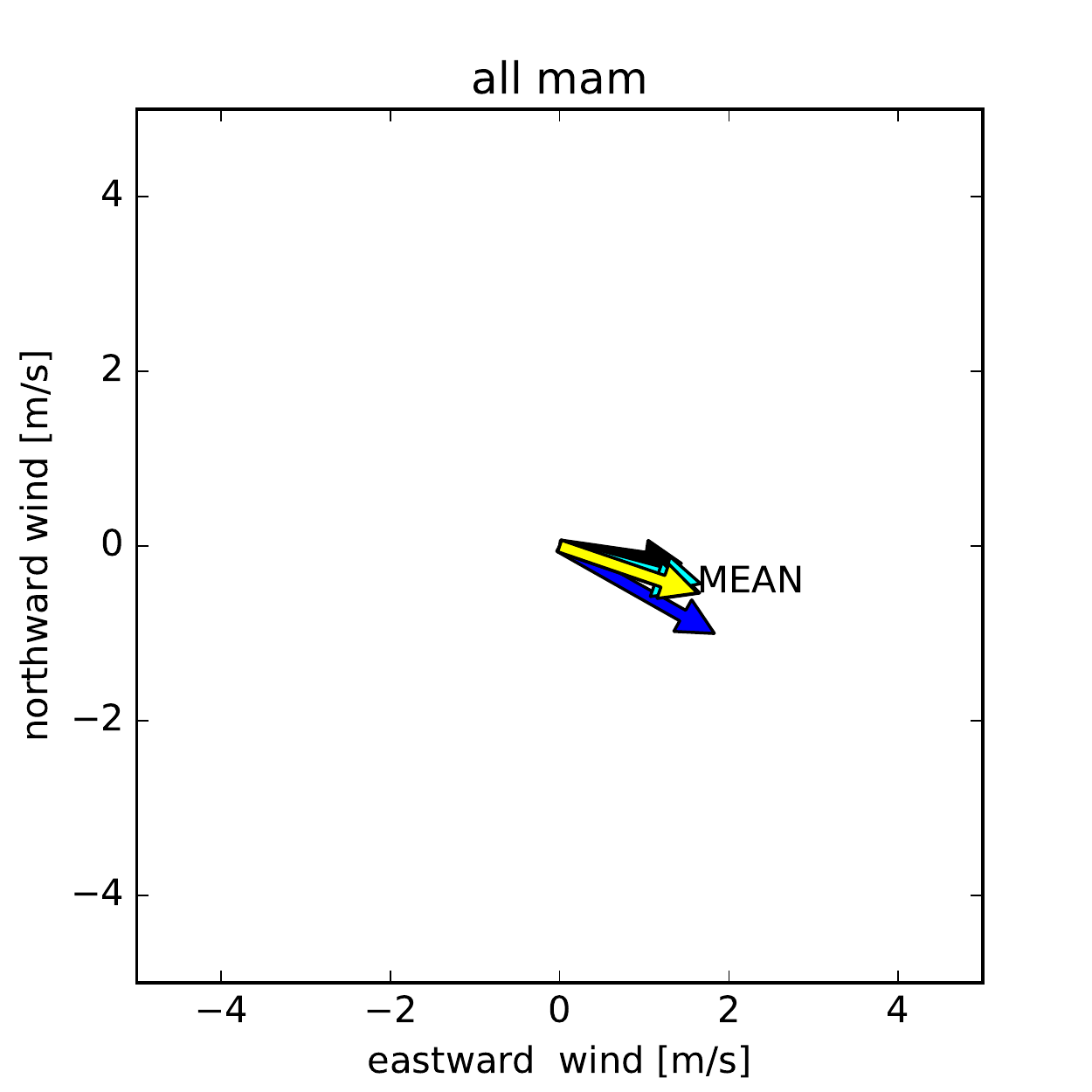}\protect}\subfloat[]{\protect\centering{}\protect\includegraphics[scale=0.25]{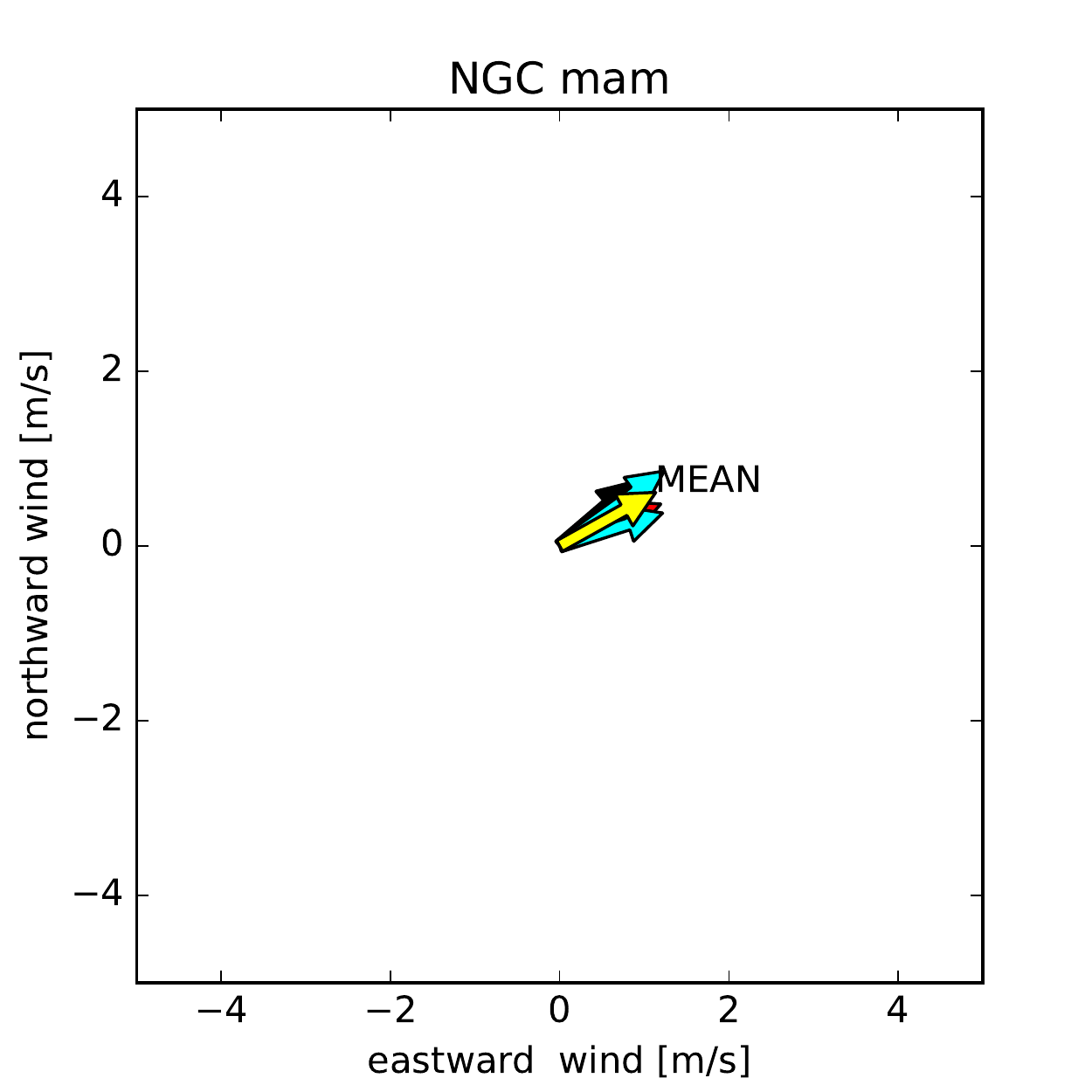}\protect}\subfloat[]{\protect\centering{}\protect\includegraphics[scale=0.25]{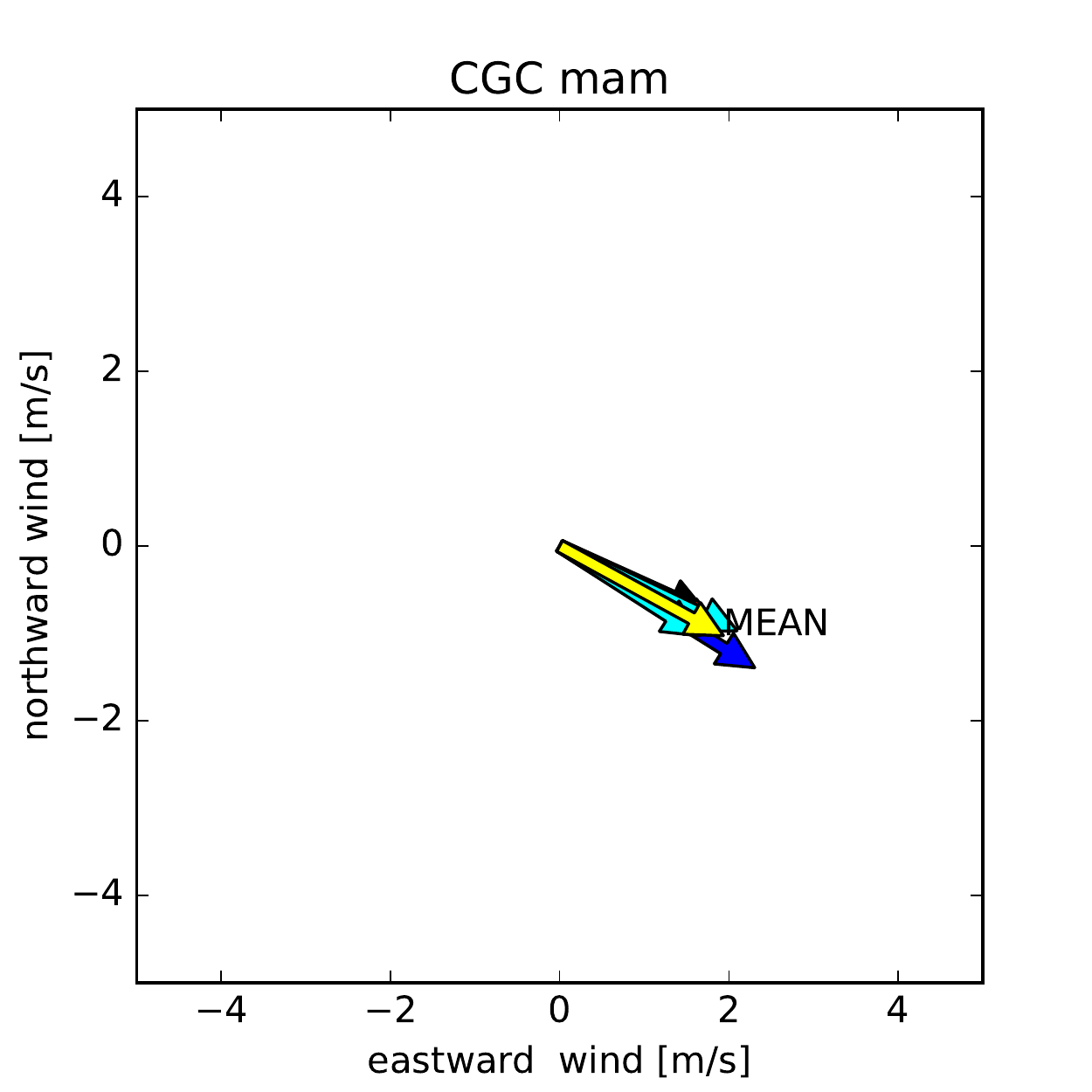}\protect}\subfloat[]{\protect\centering{}\protect\includegraphics[scale=0.25]{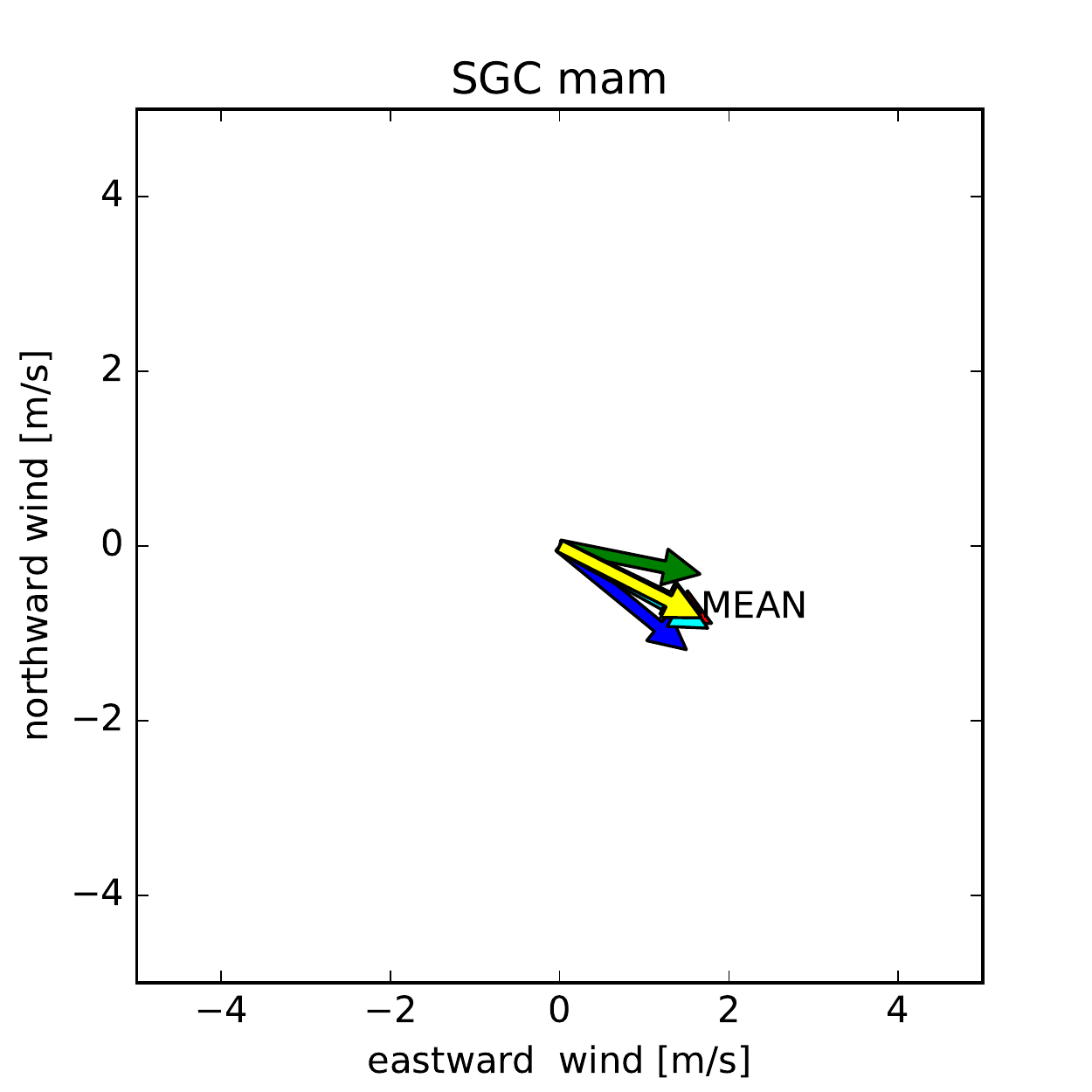}\protect}
\par\end{centering}

\begin{centering}
\subfloat[]{\protect\centering{}\protect\includegraphics[scale=0.25]{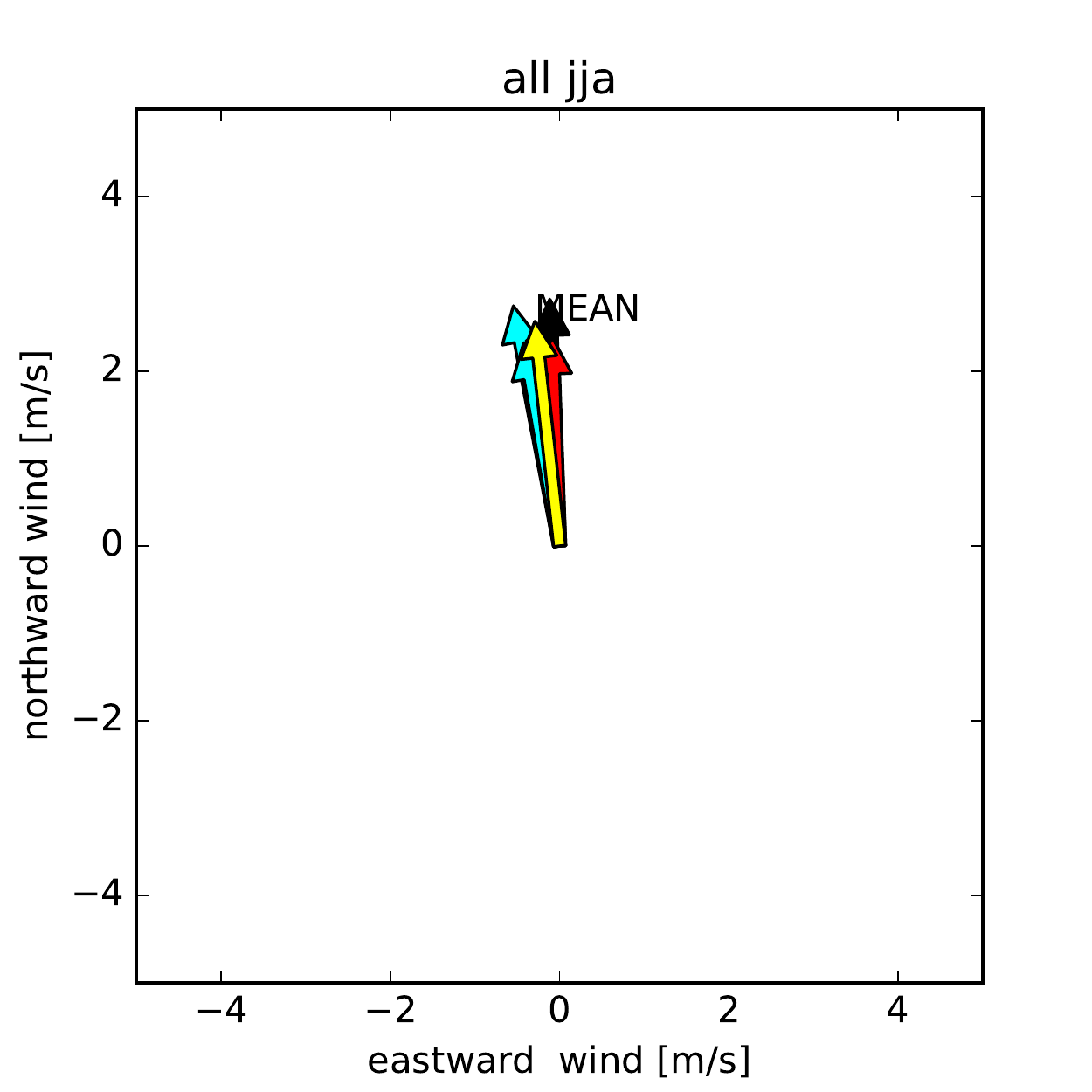}\protect}\subfloat[]{\protect\centering{}\protect\includegraphics[scale=0.25]{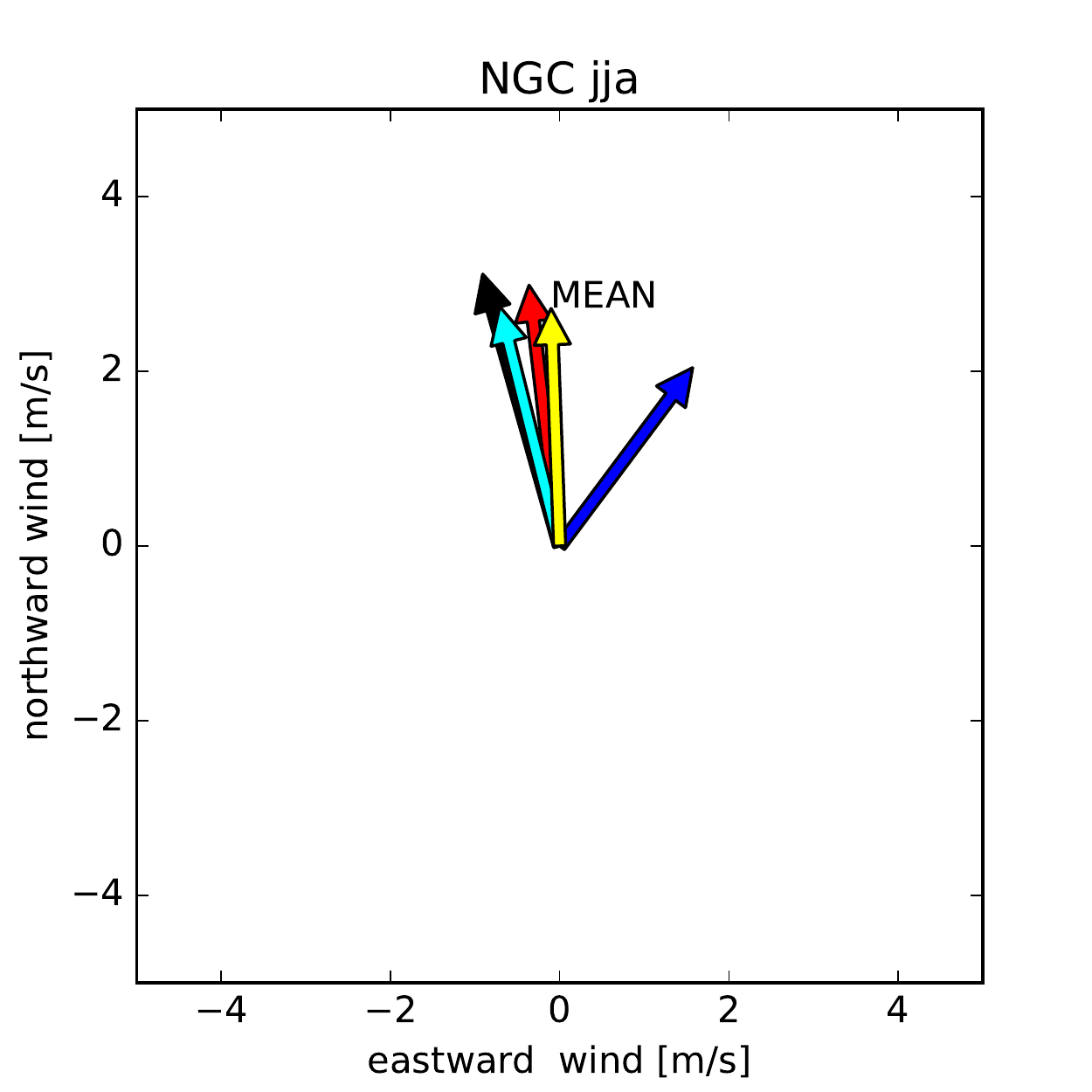}\protect}\subfloat[]{\protect\centering{}\protect\includegraphics[scale=0.25]{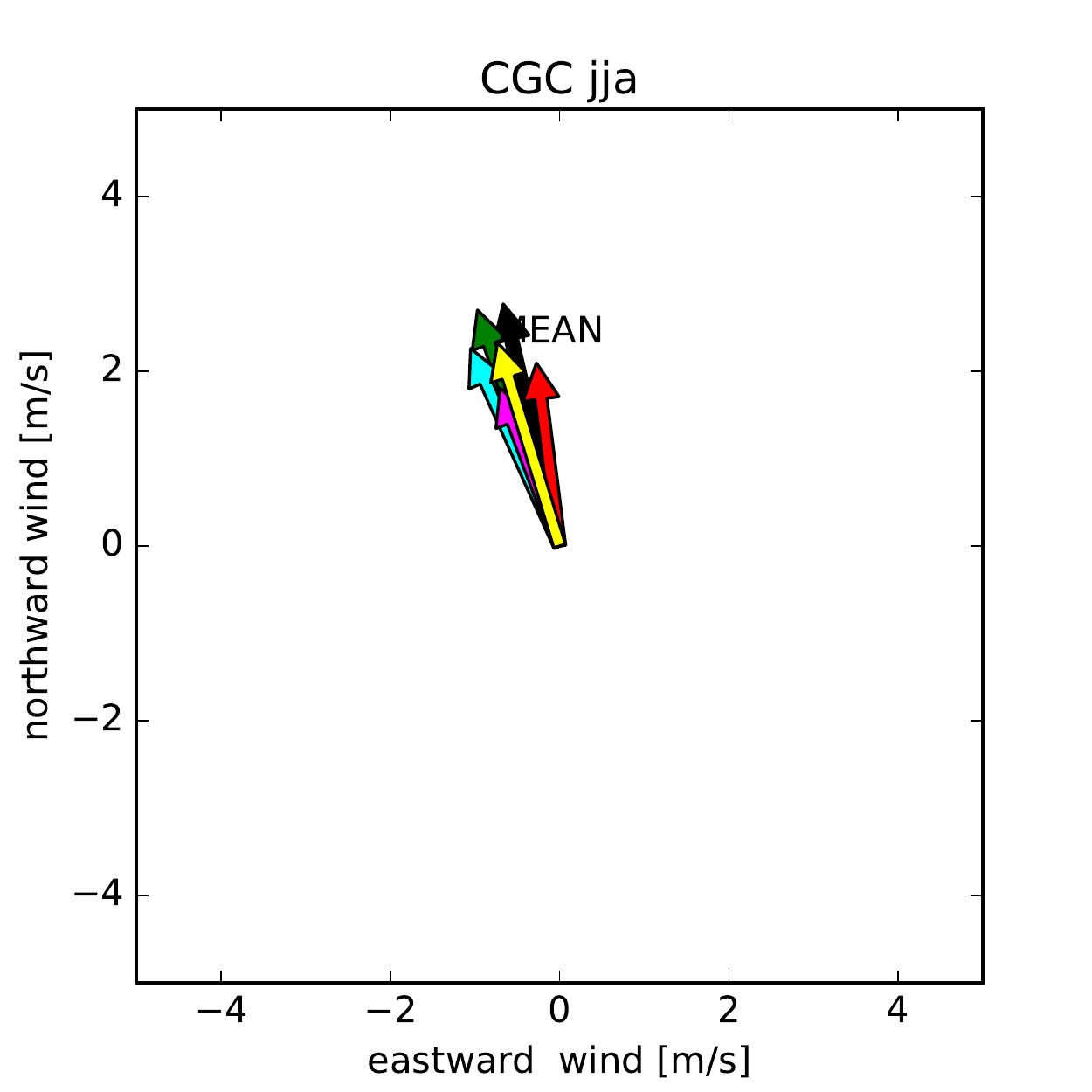}\protect}\subfloat[]{\protect\centering{}\protect\includegraphics[scale=0.25]{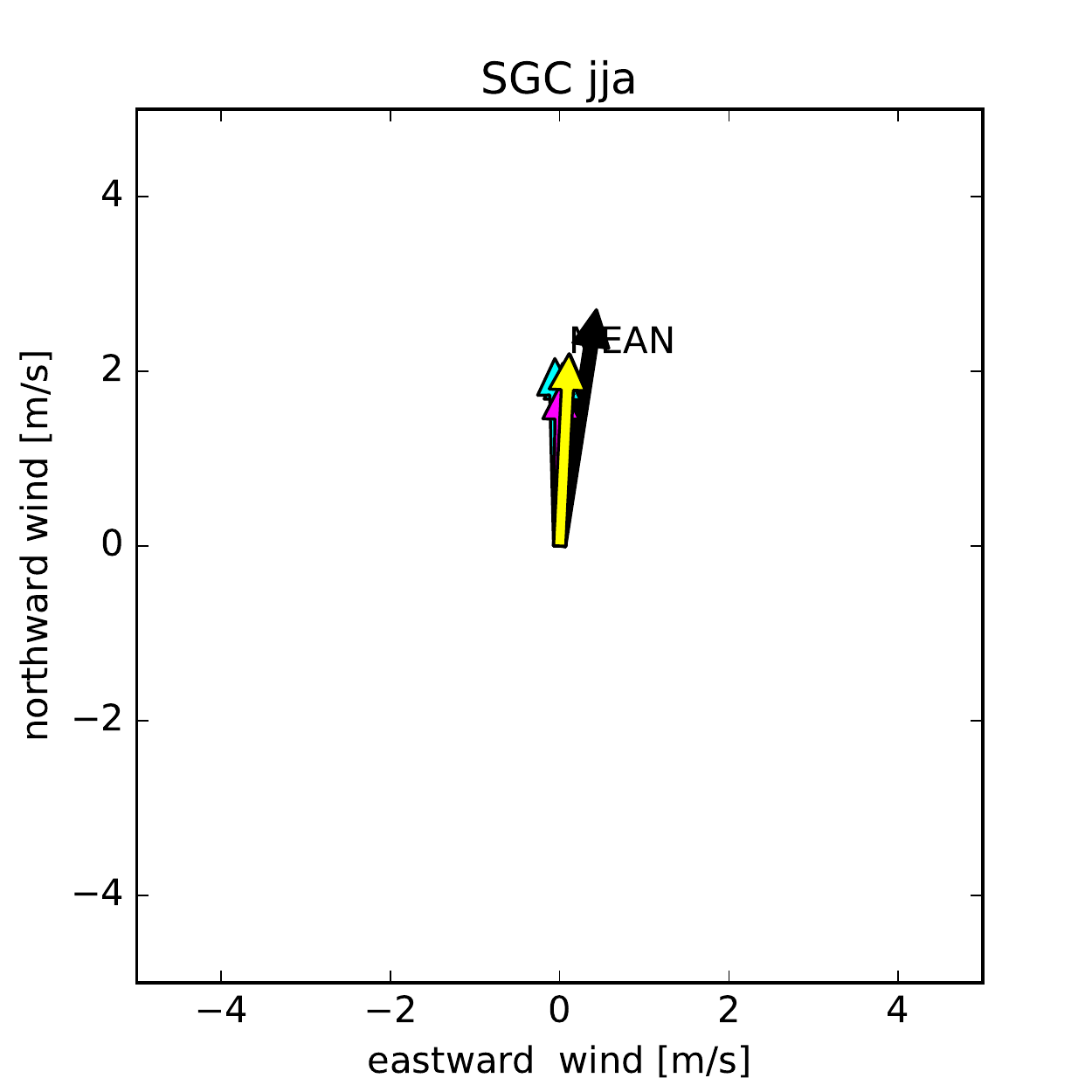}\protect}
\par\end{centering}

\begin{centering}
\subfloat[]{\protect\centering{}\protect\includegraphics[scale=0.25]{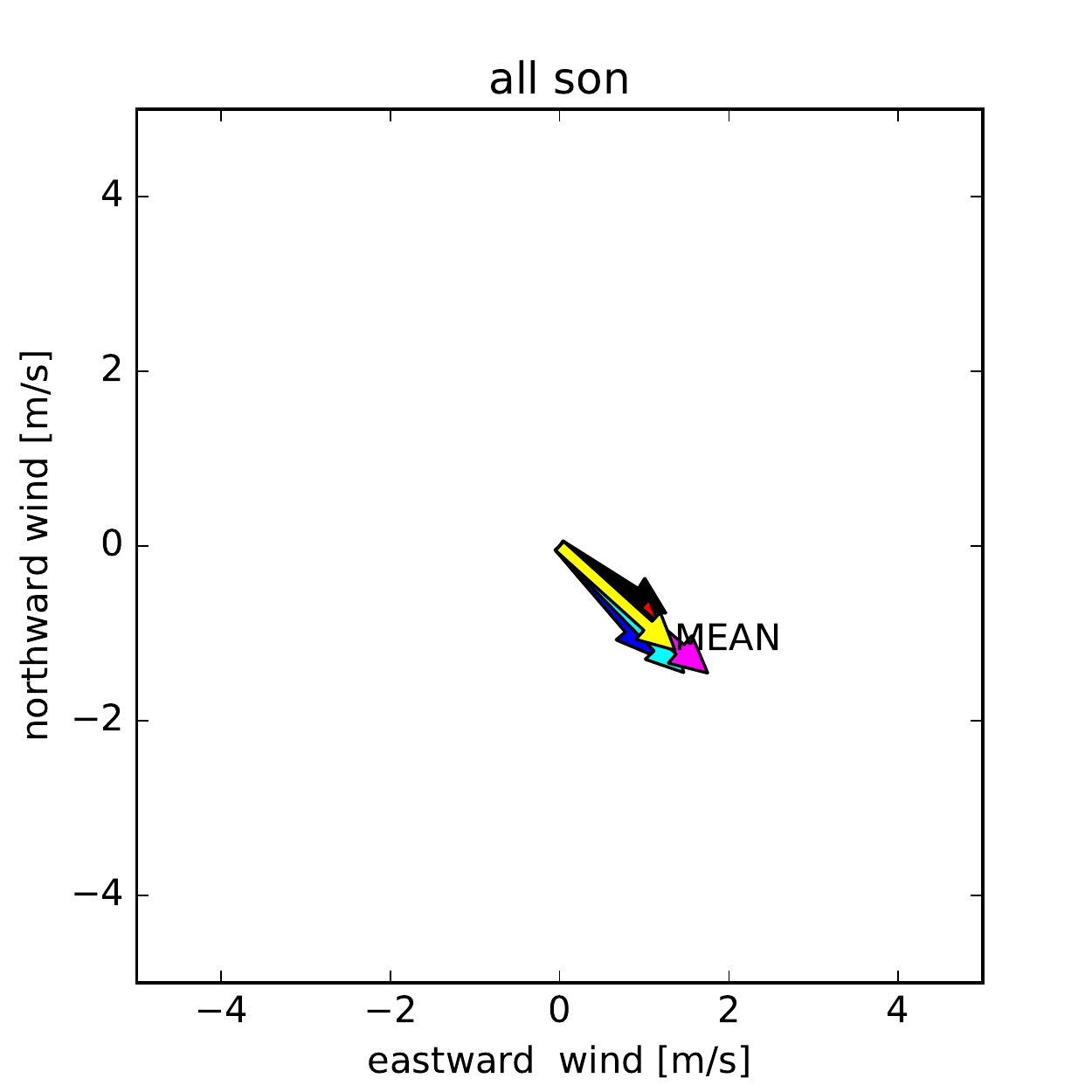}\protect}\subfloat[]{\protect\centering{}\protect\includegraphics[scale=0.25]{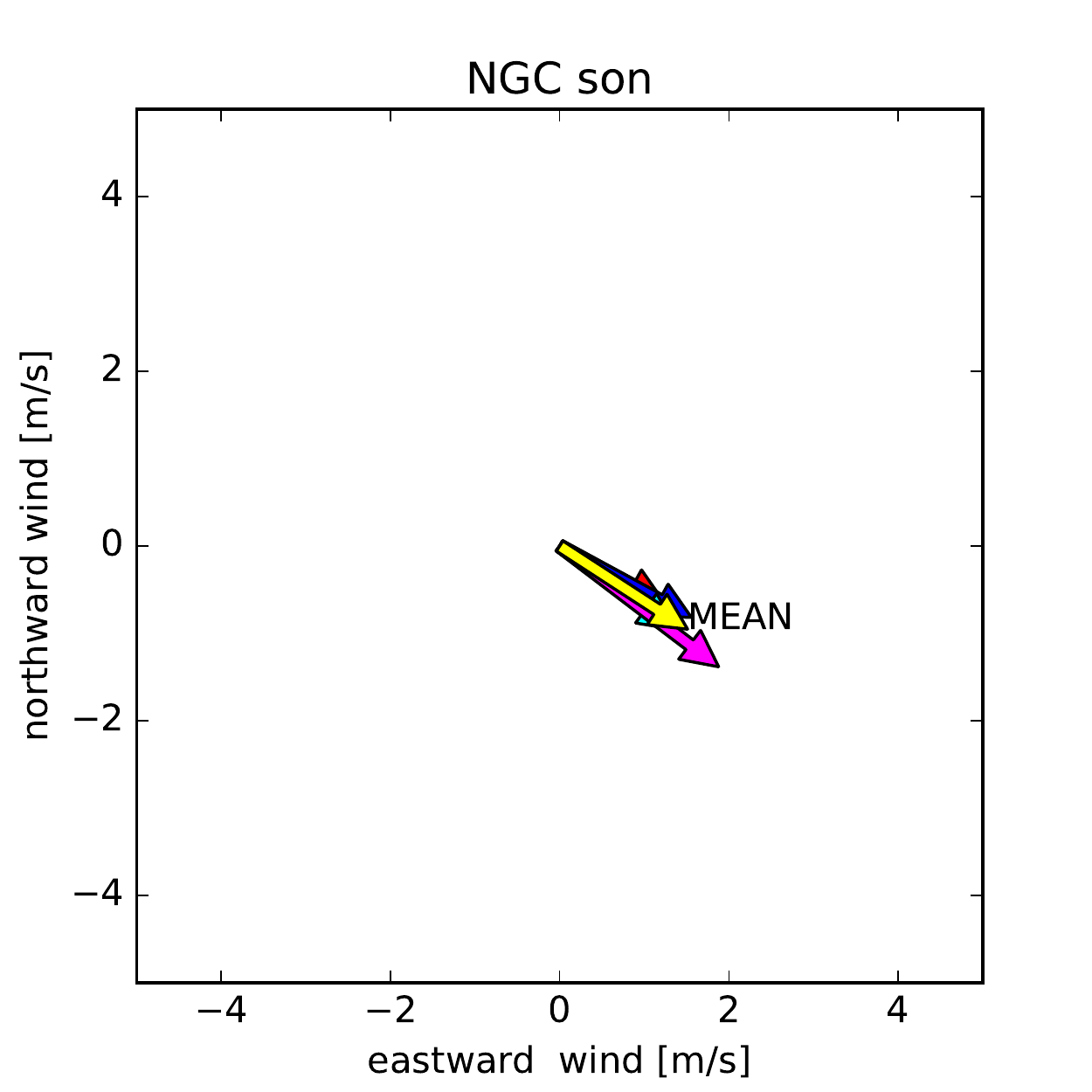}\protect}\subfloat[]{\protect\centering{}\protect\includegraphics[scale=0.25]{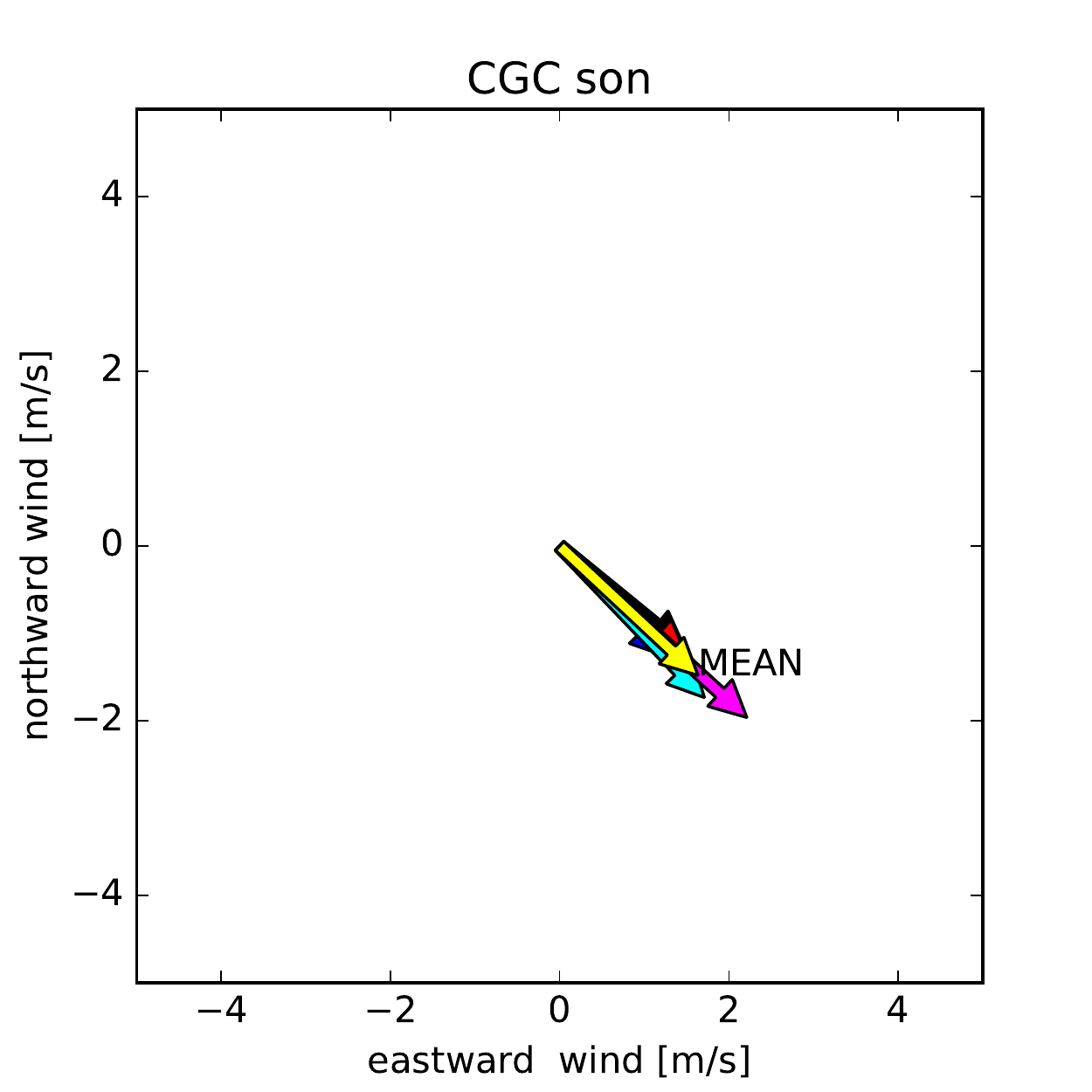}\protect}\subfloat[]{\protect\centering{}\protect\includegraphics[scale=0.25]{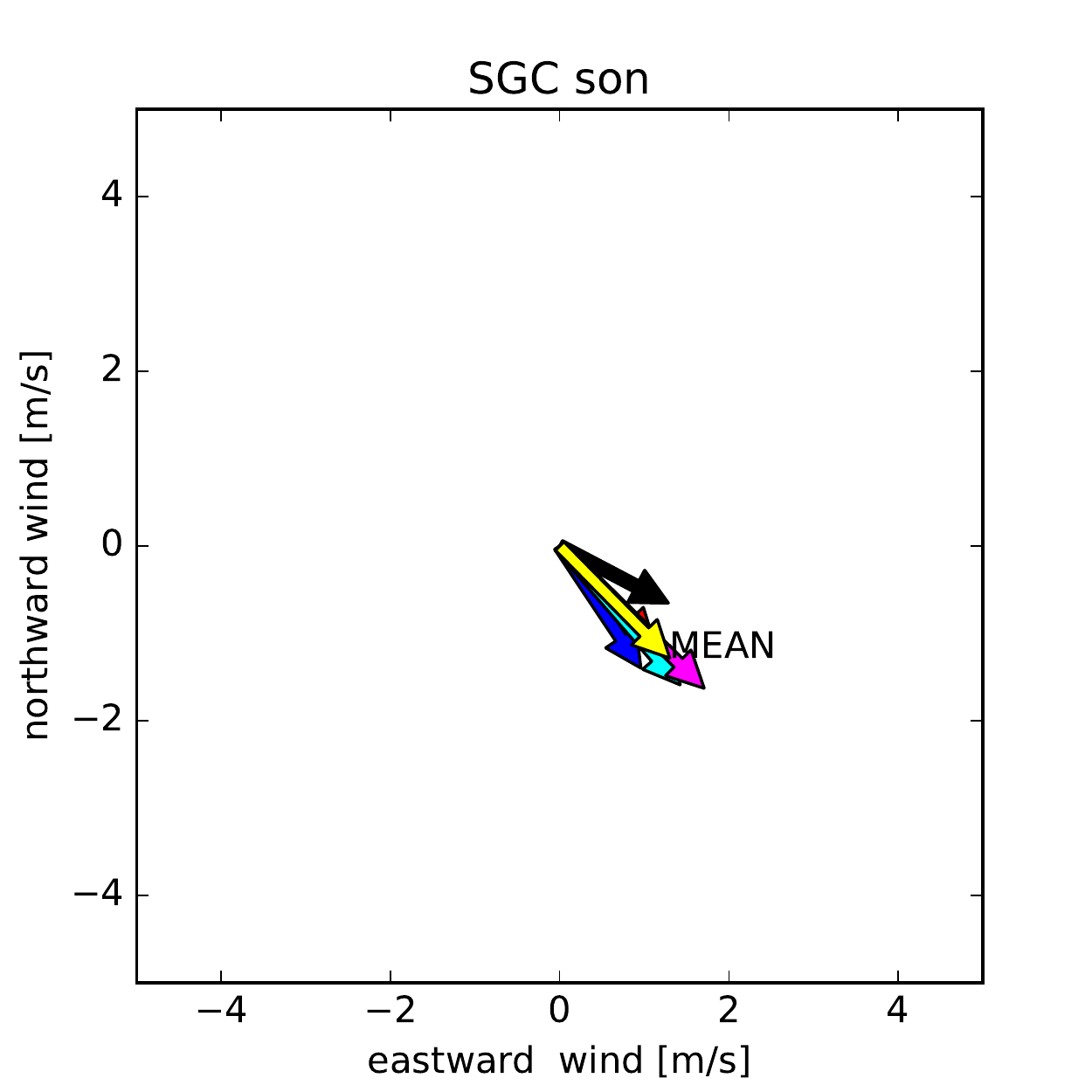}\protect}
\par\end{centering}

\centering{}\protect\caption{Temporal-spatial averages of selected dataset (same colors as in Figure~\ref{fig:Temporal-spatial-averages})
and mean (yellow)\label{fig:Temporal-spatial-averages-1}}
\end{figure}

From Figure~\ref{fig:Temporal-spatial-averages} it is clear that
whilst there is some similarity between the datasets there are also
severe outliers. With the aim of generating ``best data'' temporal-spatial
averages and recommending unique values the datasets were selected
using two criteria:
\begin{enumerate}
\item Select the datasets with a wind direction, $\alpha$, of 
\begin{equation}
\overline{\alpha}-\sigma_{\alpha}<\alpha<\overline{\alpha}+\sigma,\label{eq:crit1}
\end{equation}
where $\sigma_{\alpha}$ is the standard deviation and $\overline{\alpha}$
is the mean of the wind direction of all datasets, for the particular
sub-domain and season. This step eliminates the sever outliers apparent
in Figure~\ref{fig:Temporal-spatial-averages}.
\item From this subset datasets which have wind velocity magnitudes (wind
speeds), $v$, that are
\begin{equation}
\overline{v}-\sigma_{v}<v<\overline{v}+\sigma_{v},\label{eq:crit2}
\end{equation}
where, as before, $\sigma_{v}$ is the standard deviation and $\overline{v}$
is the mean of the wind speed across all datasets selected by criteria
one, were selected.
\end{enumerate}
The result of this selection process is shown in Table \ref{tab:selected-datased-according},
for all seasons and areas. The ``X'' entry denotes that the dataset
was retained. First according to the angle criterion, Equation~\ref{eq:crit1},
and then subsequently by the speed criterion, Equation~\ref{eq:crit2}.
The resulting means across the retained datasets are \foreignlanguage{british}{summarised}
in Table~\ref{tab:spatial-temporal-means} and illustrated graphically
in Figure~\ref{fig:Temporal-spatial-averages-1}.

\begin{table}
\centering{}%
\begin{tabular}{|c|c|c|c|c|c|c|c|c|c|c|}
\hline 
Season  & Area  & crit  & UPSCALE  & NAM  & NARR  & CCMP  & CSFR  & CERSAT V5  & CERSAT V3  & GLO \tabularnewline
\hline 
djf  & all  & angle  & X  & X  & X  & X  &  & X  & X  & X\tabularnewline
\cline{3-11} 
 &  & speed  & X  & X  &  & X  &  & X  & X  & X\tabularnewline
\cline{2-11} 
 & NGC  & angle  & X  & X  & X  & X  &  & X  & X  & X\tabularnewline
\cline{3-11} 
 &  & speed  & X  & X  &  & X  &  &  & X  & X\tabularnewline
\cline{2-11} 
 & CGC  & angle  & X  & X  & X  & X  &  & X  & X  & X\tabularnewline
\cline{3-11} 
 &  & speed  & X  & X  &  & X  &  &  & X  & X\tabularnewline
\cline{2-11} 
 & SGC  & angle  & X  & X  & X  & X  &  & X  & X  & X\tabularnewline
\cline{3-11} 
 &  & speed  & X  & X  &  & X  &  & X  &  & X\tabularnewline
\hline 
mam  & all  & angle  & X  & X  & X  & X  &  & X  & X  & X\tabularnewline
\cline{3-11} 
 &  & speed  & X  & X  &  & X  &  & X  & X  & \tabularnewline
\cline{2-11} 
 & NGC  & angle  & X  & X  &  & X  &  & X  & X  & \tabularnewline
\cline{3-11} 
 &  & speed  & X  & X  &  &  &  & X  & X & \tabularnewline
\cline{2-11} 
 & CGC  & angle  & X  & X  & X  & X  &  & X  & X  & X\tabularnewline
\cline{3-11} 
 &  & speed  & X  & X  &  & X  &  & X  & X  & \tabularnewline
\cline{2-11} 
 & SGC  & angle  & X  & X  & X  & X  &  & X  & X  & X\tabularnewline
\cline{3-11} 
 &  & speed  &  & X  & X  & X  &  & X  & X  & \tabularnewline
\hline 
jja  & all  & angle  & X  & X  & X  & X  &  & X  & X  & X\tabularnewline
\cline{3-11} 
 &  & speed  & X  & X  &  &  &  & X  & X  & \tabularnewline
\cline{2-11} 
 & NGC  & angle  & X  & X  & X  & X  &  & X  & X  & X\tabularnewline
\cline{3-11} 
 &  & speed  & X  & X  &  & X  &  &  & X  & \tabularnewline
\cline{2-11} 
 & CGC  & angle  & X  & X  & X  & X  &  & X  & X  & X\tabularnewline
\cline{3-11} 
 &  & speed  & X  & X  & X  &  &  &  & X  & X\tabularnewline
\cline{2-11} 
 & SGC  & angle  & X  & X  & X  & X  &  & X  & X  & X\tabularnewline
\cline{3-11} 
 &  & speed  & X  & X  &  &  &  & X  & X  & X\tabularnewline
\hline 
son  & all  & angle  & X  & X  &  & X  & X  &  & X  & X\tabularnewline
\cline{3-11} 
 &  & speed  & X  & X  &  & X  &  &  & X  & X\tabularnewline
\cline{2-11} 
 & NGC  & angle  & X  & X  &  & X  & X  &  & X  & X\tabularnewline
\cline{3-11} 
 &  & speed  &  & X  &  & X  &  &  & X  & X\tabularnewline
\cline{2-11} 
 & CGC  & angle  & X  & X  &  & X  & X  &  & X  & X\tabularnewline
\cline{3-11} 
 &  & speed  & X  & X  &  & X  &  &  & X  & X\tabularnewline
\cline{2-11} 
 & SGC  & angle  & X  & X  &  & X  & X  &  & X  & X\tabularnewline
\cline{3-11} 
 &  & speed  & X  & X  &  & X  &  &  & X  & X\tabularnewline
\hline 
\end{tabular}\protect\caption{Selected dataset according to the criteria (crit) Equation~\ref{eq:crit1}
(angle) and Equations~\ref{eq:crit1} and \ref{eq:crit2} (speed
subsequent to angle)\label{tab:selected-datased-according}}
\end{table}
\begin{table}
\centering{}%
\begin{tabular}{|c|c|c|c|}
\hline 
Season  & Area  & speed {[}m/s{]} & angle {[}deg{]} \tabularnewline
\hline 
djf  & all  & 4.25  & 145.88\tabularnewline
\cline{2-4} 
 & NGC  & 3.3  & 142.38\tabularnewline
\cline{2-4} 
 & CGC  & 4.62  & 141.84\tabularnewline
\cline{2-4} 
 & SGC  & 4.42  & 148.27\tabularnewline
\hline 
mam  & all  & 1.73  & 107.53\tabularnewline
\cline{2-4} 
 & NGC  & 1.31  & 61.63\tabularnewline
\cline{2-4} 
 & CGC  & 2.2  & 117.8\tabularnewline
\cline{2-4} 
 & SGC  & 1.89  & 115.74\tabularnewline
\hline 
jja  & all  & 2.6  & -6.51\tabularnewline
\cline{2-4} 
 & NGC  & 2.91  & -0.01\tabularnewline
\cline{2-4} 
 & CGC  & 2.45  & -17.39\tabularnewline
\cline{2-4} 
 & SGC  & 2.21  & 2.5\tabularnewline
\hline 
son  & all  & 1.82  & 130.9\tabularnewline
\cline{2-4} 
 & NGC  & 1.79  & 121.68\tabularnewline
\cline{2-4} 
 & CGC  & 2.21  & 132.07\tabularnewline
\cline{2-4} 
 & SGC  & 1.85  & 133.93\tabularnewline
\hline 
\end{tabular}\protect\caption{spatial temporal means\label{tab:spatial-temporal-means}}
\end{table}

\section{Dataset location and format}

The dataset is available from figshare. It comprises eight netcdf
files, containing the temporally averaged fields alongside their lat-long
coordinates. The temporal-spatial means are provided as plain ASCII
file.

\section{Dataset use and reuse}

This dataset now enables the use for forcing data in oceanographic
models, the analysis of climatological anomalies and their impacts
on the gulf dynamics and transport processes. It also allows for a
systematic analysis of the impact of wind forcing on the Gulf.

\section*{Acknowledgments}

NCEP Reanalysis data provided by the NOAA/OAR/ESRL PSD, Boulder, Colorado,
USA, from their Web site at http://www.esrl.noaa.gov/psd/.

The UPSCALE data set is licensed from the University of Reading which
includes material from NERC and the Controller of HMSO \& Queen\textquoteright s
Printer. The UPSCALE data set was created by P. L. Vidale, M. Roberts,
M. Mizielinski, J. Strachan, M.E. Demory and R. Schiemann using the
HadGEM3 model with support from NERC and the Met Office and the PRACE
Research Infrastructure resource HERMIT based in Germany at HLRS.

\bibliographystyle{plainnat}
\bibliography{lit}

\end{document}